\documentclass[lettersize,journal]{IEEEtran}

\usepackage{amsmath,amsfonts,graphics,amssymb,epsfig,subfigure,color,cite}
\usepackage{amsthm,textcomp}
\theoremstyle{plain}
\newtheoremstyle{mystyle}
  {0mm}
  {0mm}
  {}
  {4mm}
  {\bfseries}
  {:}
  { }
  {\thmname{#1}\thmnumber{ #2}\thmnote{ (#3)}}
\theoremstyle{mystyle}
\usepackage{bm}

\usepackage{array}
\usepackage{multirow}
\usepackage{enumerate}
\usepackage{algorithm}
\usepackage{algpseudocode}
\usepackage{algcompatible}
\algblockdefx{WHILE}{ENDWHILE}[1]%
  {\textbf{while }#1 \textbf{do}}%
  {\textbf{end}}
\algblockdefx{FORP}{ENDFORP}[1]%
  {\textbf{for }#1 \textbf{do in parallel}}%
  {\textbf{end}}
\algblockdefx{FOR}{ENDFOR}[1]%
  {\textbf{for }#1 \textbf{do}}%
  {\textbf{end}}
\algblockdefx{noEnd}{noEndEnd}[1]
    {#1}
    {{\Statex}}
\algblockdefx{BULLET}{EndBULLET}{$\bullet$ }{}
\algnotext{EndBULLET}
\algblockdefx{PROC}{ENDPROC}[1]
    {\textbf{procedure }#1 {}}
    {\textbf{end procedure}}
\algblockdefx{IF}{ENDIF}[1]%
  {\textbf{if }#1 \textbf{then}}%
  {\textbf{end}}
\algnewcommand\algorithmicprocedure{\textbf{procedure}}
\algnewcommand\FUNC{\item[\algorithmicprocedure]}%
\algnewcommand\algorithmicendprocedure{\textbf{end procedure}}
\algnewcommand\ENDFUNC{\item[\algorithmicendprocedure]}%
 \usepackage{setspace}
\let\Algorithm\algorithm
\renewcommand\algorithm[1][]{\Algorithm[#1]\setstretch{1.4}}
\usepackage{float}
\usepackage{makecell}
\usepackage{graphicx}
\usepackage{xcolor}
\usepackage{soul}

\newtheorem{thm}{Theorem}

\newtheorem{prop}{Proposition}

\usepackage{bbm}

\makeatletter
\newcommand{\vast}{\bBigg@{4.5}}
\newcommand{\Vast}{\bBigg@{7.5}}
\makeatother

\allowdisplaybreaks

\begin{document}
    \title{Unequal Error Protection for Digital Semantic Communication with Channel Coding}
	\author{Seonjung Kim, 
    \IEEEmembership{Student Member,~IEEE}, Yongjeong Oh, \IEEEmembership{Graduate Student Member,~IEEE}, \\ Yongjune Kim,  \IEEEmembership{Member,~IEEE}, Namyoon Lee, \IEEEmembership{Senior Member,~IEEE}, and Yo-Seb Jeon, \IEEEmembership{Member,~IEEE}
	    \thanks{Seonjung Kim, Yongjeong Oh, Yongjune Kim, Namyoon Lee, and Yo-Seb Jeon are with the Department of Electrical Engineering, POSTECH, Pohang, Gyeongbuk 37673, Republic of Korea (e-mails: \{seonjung.kim, yongjeongoh, yongjune, nylee, yoseb.jeon\}@postech.ac.kr).}
	}
	\vspace{-2mm}	
	
	\maketitle
	\vspace{-12mm}

	\begin{abstract} 
         This paper investigates unequal error protection (UEP) in digital semantic communication, where semantically important bits require substantially higher reliability than less critical ones. To characterize this heterogeneity, we introduce a novel perspective that treats learned bit-flip probabilities of semantic bits as target error protection levels, thereby directly linking semantic importance to bit-level reliability. This formulation reveals that the required protection levels of the semantic bits may differ by several orders of magnitude, making short-block coding more advantageous than conventional long-block designs. Motivated by this, we develop two UEP frameworks that minimize total blocklength under heterogeneous reliability constraints. First, we propose a bit-level UEP framework based on repetition coding, providing an analytically tractable solution that precisely meets per-bit protection requirements. Second, to improve energy and blocklength efficiency, we design a block-level UEP framework in which the semantic bits are partitioned into short blocks with similar protection levels. Guided by finite blocklength capacity analysis, we derive a closed-form threshold condition for beneficial partitioning and develop a systematic algorithm for integrating modern channel codes. Simulation results on image transmission tasks demonstrate substantial gains in both task performance and transmission efficiency compared with conventional equal-protection schemes.
    \end{abstract}


    \section{Introduction}
    Semantic communication is an emerging and disruptive paradigm that prioritizes the transmission of task-relevant information, rather than the accurate delivery of raw data bits as in traditional communication systems \cite{luo2022semantic, niu2022paradigm, park2023enabling, chaccour2024less}. \textcolor{black}{Owing to its capability to reduce both communication overhead and local processing burden, semantic communication is particularly attractive for Internet of Things (IoT) applications, where resource-constrained devices collaboratively perform remote reconstruction or inference tasks \cite{xie2020lite, yang2022semantic}.}

    A representative direction in this context is neural network-based joint source-channel coding (JSCC), which integrates source and channel coding/decoding through a neural-type encoder/decoder architecture. \textcolor{black}{Typically, the semantic encoder and decoder, placed at the transmitter (client) and receiver (server) respectively, are jointly trained in an end-to-end manner to minimize a task-specific loss function.} Neural JSCC approaches have demonstrated substantial performance gains over traditional separate source and channel coding approaches, especially in scenarios with low signal-to-noise ratio (SNR) or limited bandwidth \cite{bourtsoulatze2019deep, xie2021deep, weng2021semantic}. However, since neural encoders generally produce continuous-valued outputs, these approaches often rely on analog transmission, which lacks compatibility with modern digital communication systems and suffers from limited flexibility in rate control.  

    Digital semantic communication has gained increasing attention due to its superior compatibility and flexibility compared to the analog JSCC approach \cite{DeepJSCC_Q, yao2022semantic, zhang2024unified, NECST, park2024joint}. A representative work is \cite{DeepJSCC_Q}, where the end-to-end training of the semantic encoder and decoder was performed under finite constellation constraints (e.g., quadrature amplitude modulation), with constellation learning enabled via a differentiable soft-to-hard quantization layer. In \cite{yao2022semantic, zhang2024unified}, digital outputs were implemented by quantization of neural encoder outputs. To enable backpropagation through the quantization process, they utilized the Gumbel-softmax trick for text transmission and multi-task scenarios, respectively. For a modulation-free approach, in \cite{NECST}, semantic-related bits were generated by interpreting the neural encoder outputs as sampling probabilities, while differentiability was preserved using a low-variance gradient estimator. In \cite{park2024joint}, digital symbols were produced by sampling from encoder outputs, followed by digital modulation. A channel-adaptive modulation scheme was also introduced in \cite{park2024joint} to ensure robust task performance across varying SNRs. However, none of these methods explicitly quantify the importance of semantic features, limiting their ability to integrate with diverse channel coding and modulation schemes.

    Recent studies have attempted to quantify the importance of various types of semantic features in semantic communication \cite{sDMCM,im2024attention,park2024vision,oh2025digital,shin2026digital}.
    In \cite{sDMCM}, the inherent importance imbalance introduced by multi-level quantization was explored. Since quantization errors vary depending on the bit positions, a bit-grouping strategy was developed for modulation, treating the most significant bit (MSB) after quantization as the most important. However, this approach considers quantization error as the only source of bit-level importance and does not account for semantic relevance. In \cite{im2024attention,park2024vision}, the attention score of an image patch, extracted from a vision transformer, was used as a measure of semantic importance. Based on these scores, an importance-aware selection or quantization scheme was developed to enhance image transmission efficiency. However, such approaches are limited to image-related tasks due to their reliance on vision transformers. A universal approach was studied in \cite{oh2025digital,shin2026digital}, where the bit-flip probability of a binary symmetric channel (BSC) was interpreted as a measure of bit-level significance in semantic communication. Based on this interpretation, power allocation and modulation were optimized to meet the desired bit-flip probabilities. Despite these efforts, none of the existing methods consider channel coding design based on semantic importance, which is an essential component for ensuring reliable communication, particularly under constrained power and low SNR conditions.

    Channel coding has long been a key ingredient for addressing the unequal error protection (UEP) problem in traditional communication systems \cite{RCPC, chande2000progressive, thomos2005wireless, masnick1967linear}. For instance, in \cite{RCPC}, rate-compatible punctured convolutional codes (RCPC) were proposed to assign higher code rates to less important information blocks. However, the exact error probability analysis of convolutional codes is intractable, making real-time rate adaptation difficult in practice. In \cite{chande2000progressive, thomos2005wireless}, the assignment of code rates was optimized to different source blocks for progressive transmission using dynamic programming. While the method achieves optimality, it incurs excessive computational overhead, and estimating the expected distortion of each state remains a challenge, particularly in the context of semantic communication. In \cite{masnick1967linear}, it was shown that certain bit positions inherently possess stronger error-correction capabilities, and formal conditions were derived to characterize this behavior. Nonetheless, exploiting this potential requires maximum-likelihood decoding, thereby limiting the use of efficient bounded-distance decoding algorithms such as the Berlekamp-Massey algorithm \cite{joiner1995decoding}.

    However, the practical application of UEP has often been hindered by two main challenges. First, as discussed in \cite{RCPC, chande2000progressive, thomos2005wireless, masnick1967linear}, traditional UEP schemes incur excessive computational overhead or analytical intractability. \textcolor{black}{This issue is particularly critical in IoT scenarios, where latency, memory, and computational resources are severely constrained \cite{imteaj2021survey, samie2020hierarchical}.} Second, in conventional source data, the disparity in importance is typically not extreme enough to justify the loss of coding gain that occurs when a single long message vector is partitioned into several short message vectors \cite{bursalioglu2011unequal}. As we later establish in {\bf Theorem~1}, splitting a block is only beneficial when the difference in required protection levels between bit blocks exceeds a certain threshold. In many traditional scenarios, this condition is rarely met, making the coding gain of a long block more attractive than the granularity of UEP. However, this trade-off can be exploited beneficially in semantic communication scenarios, where the importance of task-relevant bits can vary by several orders of magnitude \cite{oh2025digital}. \textcolor{black}{Such characteristics make short-block partitioning not only beneficial for coding efficiency but also attractive for lightweight implementations in resource-constrained IoT devices.}

    Despite this potential, only a limited number of studies have investigated the UEP problem in the context of semantic communication \cite{he2023rate, fayaz2024semantic, xu2023unequal}. In \cite{he2023rate}, the authors addressed UEP for multi-modal semantic communication by applying the robust verification problem (RVP), which identifies an input feature as important if small perturbations lead to large deviations at the decoder output. In \cite{fayaz2024semantic}, the bits corresponding to segmentation maps were considered less important than those of latent maps, and the inherent UEP capability of polar codes was leveraged to provide stronger protection for the latent map bits. In \cite{xu2023unequal}, the authors utilized the attention score of each word to classify its importance into three levels, so that more important words are protected by a more robust BCH code. However, all these methods offer only modality-level, source-level or word-level error protection, rather than bit-level granularity. Addressing bit-level error protection is critical in digital semantic communication, as it enables fine-grained control over channel coding and provides greater potential for improving coding efficiency and resource utilization. To the best of the authors’ knowledge, no prior work has explored the bit-level UEP problem in digital semantic communication.

      \begin{figure*}[t]
        \centering
        {\epsfig{file=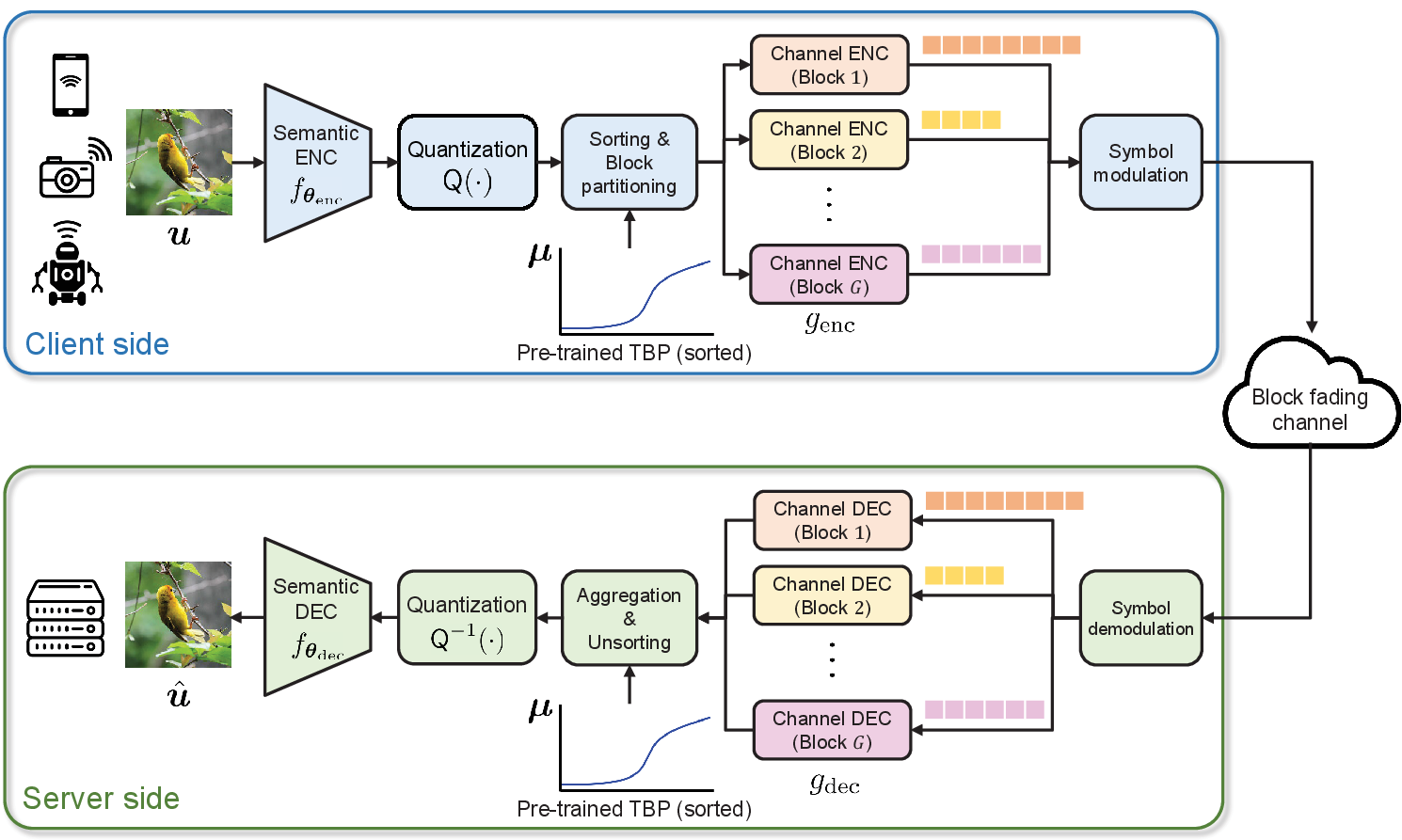,width=16cm}}\vspace{0mm}
        \caption{An illustration of a digital semantic communication system with the proposed UEP framework.}  
        \label{fig:framework}
    \end{figure*}

    To bridge this research gap, we introduce a novel perspective that treats the learned bit-flip probabilities of semantic bits as target error protection levels, thereby directly linking semantic importance to bit-level reliability. This formulation reveals that the required protection levels of semantic bits may differ by several orders of magnitude, making short-block coding more advantageous than conventional long-block designs. Inspired by this observation, we develop two UEP frameworks that minimize the total blocklength while ensuring that the actual bit-flip probability of each semantic bit does not exceed its corresponding target value.
    Simulation results demonstrate that the proposed frameworks significantly improve both task performance and transmission efficiency in digital semantic communication systems. \textcolor{black}{Furthermore, the proposed short-block UEP framework facilitates lightweight implementations for resource-constrained IoT devices and edge intelligence applications.} The main contributions of this paper are summarized as follows:
    \begin{itemize}
        \item 
        We propose a bit-level UEP framework for digital semantic communication that explicitly captures the inherent unequal error protection characteristics of semantic bits. In this framework, we formulate a bit-level UEP problem that minimizes the total blocklength while ensuring that the flip probability of each bit does not exceed its corresponding target value. The formulated problem is solved using repetition coding, where the repetition number assigned to each semantic bit is optimized to satisfy its target bit-flip probability. Leveraging the learned bit-flip probabilities of semantic bits as target protection levels provides a novel perspective that has not been explored in the existing literature.

        \item We propose a block-level UEP framework which partitions semantic bits with similar protection requirements into common blocks to exploit coding gains. Specifically, guided by finite blocklength capacity analysis, we establish that partitioning semantic bits into short blocks becomes advantageous when the difference in their required protection levels exceeds a certain threshold. By deriving this threshold in closed form, we develop a block partitioning algorithm that determines the optimal blocks for semantic bits according to their target protection levels. We further introduce a systematic method for integrating modern channel codes, such as polar and low-density parity-check (LDPC) codes, into the proposed partitioning framework. To the best of our knowledge, this is the first work to address UEP-aware block partitioning from a finite blocklength perspective in digital semantic communication systems.

    \end{itemize}

    The remainder of the paper is organized as follows. In Sec.~II, we describe the system model of a digital semantic communication scenario. In Sec.~III, we formulate a bit-level UEP problem and propose a repetition coding framework to solve this problem. In Sec.~IV, we develop a block-level UEP framework that partitions semantic bits into short blocks with similar protection levels.  In Sec.~V, we provide simulation results that validate the effectiveness of the proposed UEP frameworks across various datasets. Finally, in Sec.~VI, we present our conclusions and future research directions.

    \section{System Model}  

    In this work, we consider a wireless sensor data transmission scenario based on digital semantic communication, where a client device transmits sensor data to a server, as illustrated in Fig.~\ref{fig:framework}. For concreteness, we focus on image transmission, as it has been the most widely studied application of semantic communications. Nevertheless, the core components of the proposed framework are not limited to image data and can be readily extended to other types of sensor data.

    Given an image data $\bm{u} \in \mathbb{R}^U$, the client extracts a semantic feature from the input data using a semantic encoder as follows:
    \begin{equation}
        \bm{v} = f_{{\bm{\theta}}_{\text{enc}}}(\bm{u}) \in \mathbb{R}^M,
    \end{equation}
    where $f_{{\bm{\theta}}_{\text{enc}}}$ denotes the semantic encoder parameterized by $\bm{\theta}_{\rm enc}$, and $\bm{v}$ represents the semantic feature vector of length $M$. After encoding, each element of $\bm{v}$ is quantized using a uniform $B$-bit quantizer. The resulting quantized output is defined as
    \begin{equation}\label{eq:quantize} 
     {\bm q}_i = \textsf{Q}(v_i) \in \mathcal{Q}, \quad i \in [M], 
    \end{equation}
     where $v_i$ denotes the $i$-th element of $\bm{v}$, $\mathcal{Q} = \{\tilde{\bm q}_1, \tilde{\bm q}_2, \ldots, \tilde{\bm q}_{2^B}\}$ represents the quantizer codebook, and each $\tilde{\bm q}_i$ is a binary vector of length $B$. In this work, we refer to each binary entry of $\tilde{\bm q}_i$ as a \textit{semantic} bit. 
    The client then constructs a \textit{semantic} bit sequence as $\bm{b} = [\bm{q}_1^\top, \cdots, \bm{q}_M^\top]^\top \in \{0,1\}^{K}$, where $K = MB$ is the total number of semantic bits. To ensure robust transmission of these bits, channel coding is applied, which augments redundant bits for error correction. The encoded bit sequence is represented as
    \begin{align}
        \tilde{\bm b} = g_{\rm enc}({\bm b})\in\{0,1\}^{N_{\rm tot}},
    \end{align}
    where $g_{\rm enc}$ denotes the channel encoder. As a result of this process, the $K$ semantic bits are extended to $N_{\rm tot}$ coded bits. 

    Details of the channel encoding process considered in our work will be presented in Sec.~III and Sec.~IV.
    This encoded bitstream $\tilde{\bm b}$ is subsequently mapped to a symbol sequence $\bm{x}= [x_1, \cdots, x_T]^\top \in \mathcal{X}^T$ by digital modulation, where $\mathcal{X}$ is a constellation set and $T$ is the length of the symbol sequence.

    The received signal at time slot $t$ is given by
    \begin{align}
        \tilde{y}_t = h\sqrt{p_t}x_t + v_t, \quad t\in[T],
    \end{align}
    where $h\in\mathbb{C}$ is the complex-valued channel coefficient, $v_t \sim \mathcal{CN}(0,\sigma^2)$ denotes the additive white Gaussian noise (AWGN) with zero mean and variance $\sigma^2$, and $p_t$ is the transmission power allocated to the $t$-th symbol. \textcolor{black}{We assume perfect channel state information at the receiver and Rayleigh block fading, where $h$ remains constant over each codeword and changes independently across codewords. After perfect channel equalization, $y_t=\tilde{y}_t/h$, and the finite-blocklength analysis is applied conditionally to the resulting instantaneous effective SNR rather than to the average SNR.}


    Upon receiving the signal, the server attempts to recover a semantic bit sequence $\hat{\bm b} = [\hat{\bm q}_1^{\top},\cdots,\hat{\bm q}_M^{\top}]^{\top}\in\{0,1\}^{K}$ using channel decoding, as follows:
    \begin{align}
        \hat{\bm b} = g_{\rm dec}(\bm y),
    \end{align}
    where $g_{\rm dec}$ is the channel decoder, and ${\bm y}=[y_1,\cdots,y_T]^{\sf T}$ is a received signal vector. In our work, we assume that symbol demodulation is performed as a part of channel decoding.
    From the reconstructed bits, the server recovers each semantic feature through the dequantization process ${\sf Q}^{-1}$, as follows:
    \begin{align}
        \hat{v}_i={\sf Q}^{-1}(\hat{\bm q}_i)\in\mathbb{R}, \quad i \in[M].
    \end{align}
    The recovered feature vector is then fed into the semantic decoder to reconstruct the original input data:
    \begin{align}
        \hat{\bm u}=f_{\bm{\theta}_{\rm dec}}(\hat{\bm v})\in \mathbb{R}^{U},
    \end{align}
    where $f_{\bm{\theta}_{\rm dec}}$ denotes the semantic decoder parameterized by $\bm{\theta}_{\rm dec}$.

    \section{Bit-Level UEP Framework for Digital Semantic Communication}

    In this section, we propose a bit-level UEP framework for semantic communication. To begin, we introduce a learning-based strategy to characterize target error protection levels for semantic bits. We then formulate the bit-level UEP problem that aims at guaranteeing these protection levels. We finally present a repetition coding method to solve the formulated problem. Furthermore, the repetition numbers determined by this method will serve as the basis for the initial block partitioning in the block-level UEP framework presented in Sec.~IV.

    \subsection{Characterizing Target Error Protection Levels via BSC Modeling}\label{Sec:BlindSC}

    As described in Sec. II, each semantic bit $b_i$ undergoes channel encoding, transmission over the physical channel, and subsequent decoding at the server, resulting in $\hat{b}_i$. This end-to-end process can be equivalently represented by a channel transfer function $\mathcal{T}_i(\cdot)$, defined as
    \begin{align}
        \hat{b}_i=\mathcal{T}_i(b_i)\in \{0,1\}, \quad i\in[K],
    \end{align}
     where $\mathcal{T}_i(\cdot)$ denotes the bit-level transfer function that transforms the transmitted bit $b_i$ into the reconstructed bit $\hat{b}_i$.
    The characteristics of $\mathcal{T}_i(\cdot)$ depend on the specific digital communication configuration, including the channel coefficients, noise power, channel coding scheme, and power allocation strategy.
    Therefore, during training, accurately modeling $\mathcal{T}_i(\cdot)$ is often infeasible, not only because of the absence of precise channel knowledge, but also due to the diversity of possible communication settings (e.g., coding and power allocation schemes).

    To overcome this limitation, we adopt a simple yet effective abstraction by modeling these transfer functions as $K$ parallel BSCs as in \cite{NECST, bao2025sdac, park2024joint}. In this model, each semantic bit $b_i$ is assumed to pass through an independent\footnote{Different semantic bits may exhibit some degrees of correlation depending on the communication scenario (e.g., when block channel coding is applied). Nevertheless, we employ the independent BSC assumption solely to characterize the required bit-flip probability levels during the offline training phase. No such independence assumption is imposed during the inference phase.} BSC characterized by a unique bit-flip probability $\mu_i$. 
    \textcolor{black}{Through end-to-end training, the bit-flip probability $\mu_i$ reflects the relative error tolerance of the corresponding semantic bit. The semantic encoder learns to map semantically important features to bits associated with lower $\mu_i$, indicating a need for stronger error protection. Motivated by this relationship, we use the jointly learned bit-flip probabilities $\{\mu_i\}_{i=1}^K$ as target error protection levels during inference.}
    

    Among various approaches for determining the set of bit-flip probabilities during training, we adopt the end-to-end training method proposed in \cite{oh2025digital}. Nevertheless, it is important to note that the design of the proposed UEP schemes is universal and can be applied with any choice of bit-flip probabilities. When employing the method in  \cite{oh2025digital}, the bit-flip probabilities of the BSCs are jointly optimized with the semantic encoder and decoder, allowing the encoder to learn a mapping that assigns task-relevant features to bits associated with lower bit-flip probabilities. In other words, the framework jointly learns the feature-to-bit mapping and the corresponding bit-flip probability values. Following \cite{oh2025digital}, the set of the bit-flip probabilities, $\{\mu_i\}_{i=1}^{K}$, is jointly optimized with the semantic encoder and decoder using the following loss function:
    \begin{align}\label{eq:loss_function}
        \mathcal{L}=
        \frac{1}{U}\mathbb{E}_{\bm{u},\hat{\bm{u}}}[||\bm{u}-\hat{\bm{u}}||_2^2]
        +\frac{\lambda}{K}
        \sum_{i=1}^{K}
        \left(\frac{1}{2}-\mu_i\right)^2,
    \end{align}
    where the first term minimizes the reconstruction error, and the second term penalizes the deviations of bit-flip probabilities from the maximum bit-flip value of 0.5 with regularization weight $\lambda>0$. To enable the training of $\mu_i$ in a differentiable manner, a continuous relaxation technique is employed in \cite{oh2025digital}.

\begin{figure*}[t]
        \centering
        {\epsfig{file=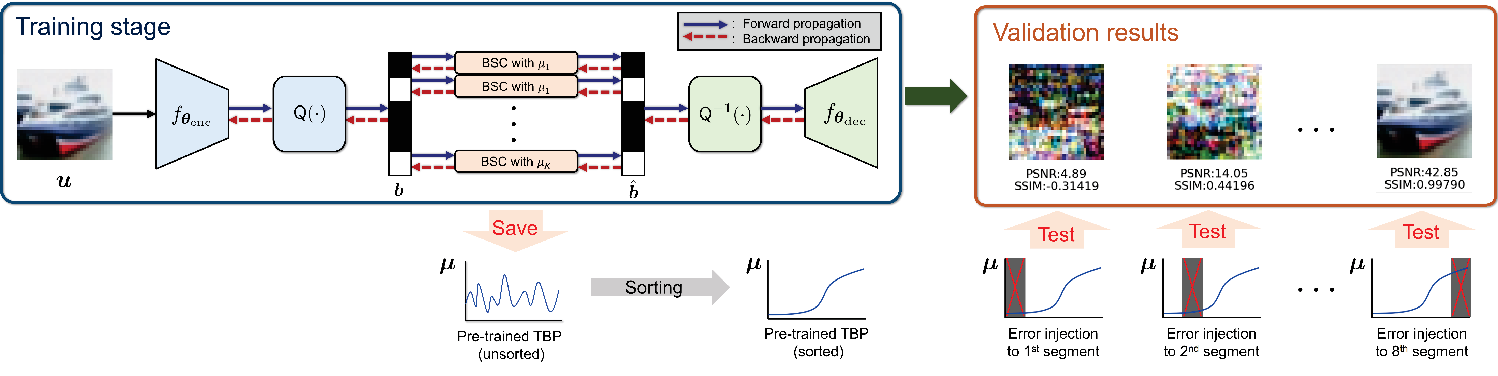,width=17cm}}\vspace{0mm}
        \caption{Visualization of reconstructed CIFAR-10 images and corresponding PSNR and SSIM values when flipping the $i$-th segment of sorted semantic bits.}  
        \label{fig:PSNR_SSIM}
    \end{figure*}

    {\bf Numerical Example (Numerical evidence for bit-flip probabilities as target protection levels):}
    In this example, we numerically demonstrate that the learned bit-flip probabilities serve as a reasonable choice for the target error protection levels. To this end, the semantic bit sequence of a test image is sorted in ascending order of the learned bit-flip probabilities. The sorted sequence is then divided into eight segments, with the first segment containing the bits with the lowest bit-flip probabilities. We flip the bits in each segment individually while keeping the remaining segments unchanged. Fig.~\ref{fig:PSNR_SSIM} visualizes the reconstructed CIFAR-10 images along with the corresponding peak signal-to-noise ratio (PSNR) and structural similarity index measure (SSIM) values when the $i$-th segment is flipped. As $i$ increases, the bits tend to have higher bit-flip probabilities, indicating lower semantic importance for the downstream task. Consequently, both PSNR and SSIM improve with increasing $i$. Notably, flipping the first segment causes a severe degradation in reconstruction quality, even resulting in a negative SSIM value, whereas flipping the eighth segment has minimal impact on the reconstructed image. These results suggest that bits with lower learned bit-flip probabilities require stronger levels of error protection.

    \subsection{Problem Formulation for Bit-Level UEP}
    
    In the bit-level UEP framework, our goal is to protect each semantic bit so that its actual bit-flip probability does not exceed the target bit-flip probability (TBP) $\mu_i$. To this end, we formulate a bit-level UEP problem, aiming to minimize the total blocklength while satisfying the TBP constraints for the semantic bits. Let ${\sf BER}(\{i\}, n_i)$ denote the bit error probability of the $i$-th semantic bit when it is encoded into $n_i$ bits. Note that ${\sf BER}(\{i\}, n_i)$ depends on both a coding rate and the channel coding scheme used. To satisfy the target bit-flip probability, ${\sf BER}(\{i\}, n_i)$ must not exceed $\mu_i$. Under these considerations, we formulate the bit-level UEP problem to minimize the total blocklength as follows:
    \begin{subequations}
    \begin{align}\label{eq:P1}
        {\textbf{(P1)}} \quad \min_{\{n_i\}_{i=1}^K} ~&
        \sum_{i=1}^K n_i \\
        \text{s.t.}~~~&
        {\sf BER}(\{i\},n_i) \leq \mu_i,~\forall i, \label{eq:P1_C1}\\
        &        n_i \in\mathbb{Z}^+, ~\forall i.\label{eq:P1_C2}
    \end{align}
    \end{subequations}

    \subsection{Simple Solution: Repetition-Coding-Based Approach}

   As a simple yet practical approach to solving the bit-level UEP problem {\bf (P1)}, we apply repetition coding to each semantic bit using bounded distance decoding. The bit error probability under bounded distance decoding with $R$ repetitions for the $i$-th semantic bit is given by
    \begin{align}
        {\sf BER}(\{i\},R)={\sf BER}_{\rm rep}(R)\triangleq\sum_{j=\lceil R/2 \rceil}^R 
        \binom{R}{j} \epsilon^j(1-\epsilon)^{R-j},
    \end{align}
    for all $i$, where ${\sf BER}_{\rm rep}(R)$ is the error probability of $R$-repetition code, and $\epsilon$ is a bit-flip probability of each coded bit. For simplicity, we assume that each semantic bit is modulated using binary phase-shift keying (BPSK) with equal power $P_{\rm trans}$. Then, the bit-flip probability of each coded bit is given by 
    $\epsilon =Q(\sqrt{2P_{\rm trans}/\sigma^2})$. It is worth noting that the above formulation is applicable to various modulation and power allocation schemes, provided that the value of $\epsilon$ is appropriately determined according to the specific modulation and power allocation strategy.

    Since an $R$-repetition code can correct up to $\lfloor (R - 1)/2 \rfloor$ symbol errors, we restrict $R$ to be an odd number, which results in a perfect code. In this case, the problem {\bf (P1)} can be decomposed into $K$ independent subproblems, each seeking the optimal number of repetitions required for a semantic bit to meet its TBP. Accordingly, we focus on minimizing the repetition number $R$ given a TBP $\mu_i$. This minimization problem can be expressed as
    \begin{subequations}
    \begin{align}\label{eq:P1_prime}
        {(\textbf{P1}^{\prime})} \quad &\min_{r_i} \quad
        R_i\\
        & ~~\text{s.t.} 
        \quad 
        {\sf BER}_{\rm rep}(R_i) \leq \mu_i,\label{eq:P1_prime_C1}\\
        &\hphantom{~~ \text{s.t.}  \quad}
        R_i=2r_i+1,\label{eq:P1_prime_C2}\\
        &\hphantom{~~ \text{s.t.}  \quad}
        r_i\in\mathbb{Z}^+,\label{eq:P1_prime_C3}
    \end{align}
    \end{subequations}
    where the constraints \eqref{eq:P1_prime_C2} and \eqref{eq:P1_prime_C3} enforce the repetition number to be an odd number.

    Since ${\sf BER}_{\rm rep}(2r+1)$ is strictly decreasing in $r$, the minimum value of $r\in\mathbb{Z}^+$ which satisfies ${\sf BER}_{\rm rep}(2r+1)\leq \mu$ can be determined using a bisection method. 
    To establish the initial points for the bisection method, we set an upper bound of $r$ as
    \begin{align}\label{eq:initial_points}
        \textcolor{black}{r^{\rm (ub)}=\left\lceil
        \frac{1}{2}\left(\frac{\ln \mu}{\ln(2\sqrt{(\epsilon(1-\epsilon)})} -1\right)
        \right\rceil.}
    \end{align}
    The proof of the boundedness of $r^{\rm (ub)}$ is given in the following proposition:

    \begin{prop}\label{prop:bound_of_bisection}
        Suppose that $r^{\rm (ub)}$ is defined as in \eqref{eq:initial_points}, given a TBP $\mu$ and a coded-bit-flip probability $\epsilon$. Then, the following inequality holds:
        \begin{align}\label{eq:optimal_rate}
            {\sf BER}_{\rm rep}(2r^{\rm (ub)}+1) \leq \mu.
        \end{align}
    \end{prop}

    \begin{IEEEproof}
        See Appendix~\ref{Apdx:Prop1}.
    \end{IEEEproof}
    Additionally, to account for the fact that $P_{\rm rep}(2r+1)$ can only be evaluated at integer values, we modify the bisection method to track both the ceiling and floor values of the midpoint during the search. The detailed procedure is described in {\bf Algorithm~1}.

    \begin{algorithm}[t]
        \caption{Bisection method for determining the required repetition number}\label{alg:Bisection}
        {\small \begin{algorithmic}[1]
            \REQUIRE $\mu_i\text{ and } \epsilon$ 
            \STATE  \textbf{Initialize:} $r^{\rm (lb)} \gets 0, \textcolor{black}{r^{\rm (ub)} \gets \left\lceil
            \frac{1}{2}\left(\frac{\ln \mu_i}{\ln(2\sqrt{(\epsilon(1-\epsilon)})} -1\right)
            \right\rceil}$
            \IF {$\epsilon<\mu_i$}
            \STATE $r_i \gets 0$
            \STATE \textbf{Break}
            \ELSE
            \REPEAT
            \STATE $c^{\rm (lb)}\gets \left\lfloor
            \frac{r^{\rm (lb)}+r^{\rm (ub)}}{2}
            \right\rfloor, c^{\rm (ub)}\gets \left\lceil
            \frac{r^{\rm (lb)}+r^{\rm (ub)}}{2}
            \right\rceil$
            \IF {$P_{\rm rep}(2c^{\rm (ub)}+1)>\mu_i$}
            \STATE $r^{\rm (lb)}\gets c^{\rm (ub)}$
            \ELSIF {$P_{\rm rep}(2c^{\rm (lb)}+1)<\mu_i$}
            \STATE $r^{\rm (ub)}\gets c^{\rm (lb)}$
            \ELSE
            \STATE $r_i \gets c^{\rm (ub)}$
            \STATE \textbf{Break}
            \ENDIF
            \UNTIL
            \ENDIF
            \STATE $R_i=2r_i+1$
            \ENSURE $R_i$
        \end{algorithmic}}
    \end{algorithm}

    Despite the computational efficiency of the bisection method, applying it independently to each bit incurs excessive complexity proportional to $K$. To mitigate this complexity in determining the repetition numbers for all semantic bits, we develop a low-complexity approach for computing $\{R_i\}_{i=1}^K$. 
    In this approach, we first sort\footnote{In practice, this sorting necessitates corresponding reordering and inverse reordering at the client and server, respectively, as illustrated in Fig.~1. This operation incurs no additional communication overhead since the learned bit-flip probabilities are shared between the client and server during offline training, making this assumption practical.} the semantic bits $\bm{b}$ in ascending order based on their TBP ${\mu_i}$.
    Then, we leverage the ordering of the TBP $\{\mu_i\}_{i=1}^K$, which induces a monotonic constraint on $\{R_i\}_{i=1}^K$, i.e., $R_i \geq R_{i+1}$ for all $i$. Specifically, we determine the maximum repetition number, $R_1$, by applying the bisection method to the semantic bit with the most stringent requirement, $\mu_1$. Subsequently, we identify the transition points at which the repetition number changes, which occur when
    \begin{align}
        \mu_{i} < \mu_{i+1} \Rightarrow R_{i} > R_{i+1}.
    \end{align}
    This approach requires only a single invocation of the bisection method, thereby significantly reducing the computational complexity compared to the na\"{i}ve per-bit optimization strategy given in {\bf Algorithm~1}. The detailed procedure is described in {\bf Algorithm~2}.
 

    {\color{black}
    \begin{algorithm}[t]
        \color{black}
        \caption{Low complexity algorithm for determining the required repetition numbers}\label{alg:low_complexity}
        {\small \begin{algorithmic}[1]
            \REQUIRE $\{\mu_i\}_{i=1}^K\text{ and } \epsilon$
            \STATE $R \gets \textbf{Algorithm 1}(\mu_1, \epsilon)$
            \IF {$R>1$}
                \STATE $\mu^{\rm (tr)} \gets P_{\rm rep}(R-2)$
            \ENDIF
            \FOR {$i=1$ to $K$}
                \WHILE {$R>1$}
                    \IF {$\mu^{\rm (tr)}\leq \mu_i$}
                        \STATE $R \gets R-2$
                        \IF {$R>1$}
                            \STATE $\mu^{\rm (tr)} \gets P_{\rm rep}(R-2)$
                        \ENDIF
                    \ELSE
                        \STATE \textbf{Break}
                    \ENDIF
                \ENDWHILE
                \STATE $R_i \gets R$
            \ENDFOR
            \ENSURE $\{R_i\}_{i=1}^K$
        \end{algorithmic}}
    \end{algorithm}}

    \section{Block-Level UEP Framework for Digital Semantic Communication}
    Although repetition coding offers analytical tractability for satisfying the bit-level UEP constraint, it provides no channel coding gain and exhibits poor transmission efficiency compared to modern channel codes. In semantic communication based on the method in Sec.~\ref{Sec:BlindSC}, certain semantic bits may share similar TBP. This observation suggests that applying block coding to such blocks of bits can yield improved coding gain without significantly violating the UEP constraint. Motivated by this insight, we propose a \textit{block-level} UEP framework that enables partitioning of bits with similar TBP based on the finite blocklength capacity analysis.

    \subsection{Problem Formulation for Block-Level UEP}
    In the proposed block-level UEP framework, we partition the $K$ semantic bits into $G$ blocks based on their TBP. Let $\mathcal{K}_i$ denote the set of indices corresponding to the $i$-th block, consisting of $k_i$ bits, given by 
    \begin{align}
        \mathcal{K}_i = \{\bar{k}_{i-1}+1, \bar{k}_{i-1}+2, \ldots, \bar{k}_{i-1} +k_i \},
    \end{align}
    where $\bar{k}_{i-1} = \sum_{j=1}^{i-1}k_j$.
    We then jointly encode the semantic bits within each block using a block channel encoder, without explicitly applying the UEP principle. Instead, the encoder is designed to satisfy the most stringent protection level within the block, i.e., to support the minimum TBP among the bits in the block. Specifically, an $(n_i, k_i)$ block code is applied to the semantic bits in block $\mathcal{K}_i$ such that the resulting bit error probability, denoted by $P_{\rm ib}(\mathcal{K}_i, n_i)$, satisfies the following inequality:
    \begin{align}
        P_{\rm ib}(\mathcal{K}_i,n_i)\leq \min_{j\in \mathcal{K}_i} \; \mu_j, \; \forall i\in[G].
    \end{align}
    Note that the allowable values of $(n_i, k_i)$ are constrained by the structure of the specific coding scheme\footnote{For example, when employing a binary primitive BCH code with error-correcting capability 2 and codeword length $n_i \leq 255$, the valid $(n_i, k_i)$ pairs are given by $\{(15,7),(31,21),(63,51),(127,113),(255,239)\}$ where $n_i$ is determined by $k_i$ \cite{lin_costello}.}.

    Considering the constraints on block partitioning and coding rate, we formulate the block-level UEP problem as follows:

    \begin{subequations}
    \begin{align}\label{eq:P2}
        {\textbf{(P2)}}~~\min_{\{\mathcal{K}_i, n_i\}_{i=1}^G, G} ~~&
        \sum_{i=1}^G n_i\\
         \text{s.t.}~~~~~~~& 
         P_{\rm ib}(\mathcal{K}_i, n_i)\leq \min_{j\in \mathcal{K}_i}\mu_j,\; \forall i\in[G],\label{eq:P2_C1}\\
        &\mathcal{K}_{i}\cap \mathcal{K}_{i^\prime}=\emptyset,\;\forall i\neq i^\prime,\label{eq:P2_C2}\\
        &\cup_i \mathcal{K}_i = [K],\label{eq:P2_C3}\\
        &|\mathcal{K}_i|\in \overline{\mathcal{K}},\; \forall i\in[G],\label{eq:P2_C4}\\
        &n_i\in \mathcal{N}(|\mathcal{K}_i|),\; \forall i \in[G], \label{eq:P2_C5}\\
        &G\in\mathbb{Z}^+,
        \label{eq:P2_C6}
    \end{align}
    \end{subequations}
    where $\overline{\mathcal{K}}$ is the set of possible block sizes before encoding, and $\mathcal{N}(|\mathcal{K}_i|)$ is the set of possible block sizes after encoding given $|\mathcal{K}_i|$. In the above problem, the constraint \eqref{eq:P2_C2} makes all blocks disjoint from each other, and \eqref{eq:P2_C3} makes the blocks cover all bits. Note that the constraints on possible block sizes, $\mathcal{K}$ and $\mathcal{N}(|\mathcal{K}_i|)$, depend on the channel coding scheme applied.

    \subsection{Theoretical Foundation for Block Partitioning}
    As seen in the block-level UEP problem {\bf (P2)}, proper partitioning of the semantic bits is critical in maximizing the performance of the block coding approach. For example, when bits have similar TBP, forming long blocks can reduce the overall blocklength and improve coding efficiency. In contrast, when the target probabilities differ significantly, using short blocks allows for more precise adherence to individual protection requirements, thereby enhancing overall UEP performance. We formalize this intuition by using the normal approximation of the finite blocklength capacity, which approximates the rate $k/n$ of an $(n,k)$ code as
    \begin{align}
        \frac{k}{n}\approx \log_2(1+\textsf{SNR})-\sqrt{\frac{V}{n}}\frac{Q^{-1}(\textsf{BLER})}{\ln 2},
    \end{align}
    where \textsf{BLER} is the block error rate (BLER) of an $(n,k)$ code and $V=1-(1+\textsf{SNR})^{-2}$ is the channel dispersion \cite{polyanskiy2010channel}. Based on this, we approximate a BLER expression for an $(n,k)$ code as follows:
    \begin{align}\label{eq:BLER}
        \textsf{BLER}(n,k)=Q(f(n,k)),
    \end{align}
    where $f(n,k)=\ln 2\sqrt{\frac{n}{V}}(\log_2(1+\textsf{SNR})-\frac{k}{n})$ \cite{ren2019joint}. 

    On the basis of the BLER expression given in \eqref{eq:BLER}, we prove that a blocklength can be reduced by partitioning semantic bits into blocks of the same protection level. This result is formally stated in the following theorem:
    \vspace{2mm}
    \begin{prop}\label{prop:motivation_1}
    Consider two block codes: an $(n_1,k_1)$ code and an $(n_2,k_2)$ code, whose BLERs are given as in \eqref{eq:BLER}.
    If $\textsf{BLER}(n_1,k_1)=\textsf{BLER}(n_2,k_2)$, there exists a positive integer $n_3<n_1+n_2$ such that 
    \begin{align}
        \textsf{BLER}(n_3,k_1+k_2) = \textsf{BLER}(n_1,k_1).
    \end{align}
    \end{prop}

    \begin{IEEEproof}	
    	See Appendix~\ref{Apdx:Thm1}.	
    \end{IEEEproof}
    \vspace{2mm}

    Additionally, we formally characterize the condition under which block partitioning results in a longer blocklength. This finding is stated in the following theorem:
    \vspace{2mm}
    \begin{thm}\label{thm:motivation_2}
    Consider three block codes: an $(n_1,k_1)$ code, an $(n_2,k_2)$ code, and an $(n_3,k_1+k_2)$ code, whose BLERs are given as in \eqref{eq:BLER}. 
    If $\textsf{BLER}(n_3,k_1+k_2) = \textsf{BLER}(n_1,k_1)$, 
    then a necessary and sufficient condition for $n_3>n_1+n_2$ is
    \begin{align}\label{eq:split_condition}
        \frac{Q^{-1}(\textsf{BLER}(n_1,k_1))}{Q^{-1}(\textsf{BLER}(n_2,k_2))} \geq \gamma_{\rm th}(n_1,n_2,k_1,k_2),
    \end{align}
    provided that $\gamma_{\rm th}(n_1,n_2,k_1,k_2)$ is given by
    \begin{align}\label{eq:C_star}
        &\gamma_{\rm th}(n_1,n_2,k_1,k_2) \nonumber \\
        &=\frac{\sqrt{2Y(\sqrt{Y^2+4(k_1+k_2)C}-\sqrt{Y^2+4k_1C})+4k_2C}}{\sqrt{Y^2+4(k_1+k_2)C}-\sqrt{Y^2+4k_1C}},
    \end{align}
    where $C=\log_2(1+\textsf{SNR})$, and $Y=\sqrt{n_1}C-k_1/\sqrt{n_1}$.
    \end{thm}
    \begin{IEEEproof}	
    	See Appendix~\ref{Apdx:Thm2}.	
    \end{IEEEproof}
    \vspace{2mm}

    \subsection{Proposed Block Partitioning Algorithm}
    \textbf{Proposition~2} establishes the advantage of partitioning semantic bits with similar target protection levels into common blocks. Conversely, \textbf{Theorem~1} emphasizes the necessity of splitting blocks when semantic bits have significantly different protection levels. The left-hand side of \eqref{eq:split_condition} quantifies the BLER difference between two blocks. Specifically, as $\textsf{BLER}(n_2,k_2)$ increases, the ratio $Q^{-1}(\textsf{BLER}(n_1,k_1))/Q^{-1}(\textsf{BLER}(n_2,k_2))$ also increases, making separation of the blocks preferable. This intuitive result provides a quantitative criterion for deciding when to merge or separate blocks. Motivated by this observation, we propose a block partitioning algorithm that minimizes the overall blocklength while ensuring that the TBP are satisfied.

    \textbf{Proposition~2} indicates that partitioning comparable error protection requirements into common blocks reduces blocklength, but it does not explicitly define what constitutes \textit{comparable}. To address this, we leverage a repetition-based UEP criterion: each semantic bit $i$ is assigned a repetition number $R_i$ such that its bit-flip probability remains below the target $\mu_i$. Bits sharing the same repetition number inherently possess equivalent protection levels, as they receive identical error protection. We thus partition bits according to identical $R_i$ values, exploiting coding gain without violating repetition-based UEP constraints. Specifically, the first block includes bits whose repetition numbers are $R_{\max}$, the second block includes bits with repetition number $R_{\max}-2$, and so forth, until the final block contains bits with the minimum repetition number (i.e., $R_i=3$).
    It is worth noting that block channel coding is not applied to semantic bits with a repetition number of one (i.e., $R_i = 1$), as these bits already meet their target error protection requirements without the need for additional coding.

    After this initial partitioning, the total number of the non-singleton bit blocks is $(\max_i R_i -1)/2$. To further reduce the overall blocklength, we utilize the criterion derived from \textbf{Theorem~1}. Although \textbf{Theorem~1} originally provides an equivalent condition favoring separation, it implicitly produces the following condition for merging blocks:
   \begin{align}
        \frac{Q^{-1}(\textsf{BLER}(n_1,k_1))}{Q^{-1}(\textsf{BLER}(n_2,k_2))} < \gamma_{\rm th}(n_1,n_2,k_1,k_2).
    \end{align}
    However, since the exact blocklengths $n_1$ and $n_2$ required to achieve the TBP for blocks of sizes $k_1$ and $k_2$ respectively are unknown, we approximate the BLER of any block $\mathcal{A}$ as follows:
     \begin{align}\label{eq:approximated_BLER}
        \textsf{BLER}(\mathcal{A})\approx 1-\Big(1-\min_{i\in \mathcal{A}}\mu_i\Big)^{|\mathcal{A}|},
    \end{align}
    which estimates the probability that at least one bit within a block of size $|\mathcal{A}|$ is erroneous \cite{olmos2009exponential}. \textcolor{black}{The proposed block partitioning algorithm is not restricted to this particular approximation; any more accurate code- or decoder-specific BLER relation or empirical lookup table can replace \eqref{eq:approximated_BLER} without changing the overall procedure. We adopt \eqref{eq:approximated_BLER} as a simple and analytically tractable instance, and our simulations with practical polar and LDPC codes demonstrate performance gains over the corresponding EEP schemes.} 
    
    \textcolor{black}{To maintain the operating domain of \textbf{Theorem~1}, its partitioning condition is applied only when $0<\textsf{BLER}<0.5$. Whenever the approximation in \eqref{eq:approximated_BLER} exceeds $0.5$, it is clipped to $0.5$ before being used in \textbf{Algorithm~3}.} Using this approximation, we further obtain 
    \begin{align}
        Y\approx \frac{\sqrt{V}}{\ln 2}Q^{-1}(\textsf{BLER}(\mathcal{A})),
    \end{align}
    allowing us to apply the merging condition effectively. The proposed block partitioning algorithm is detailed in \textbf{Algorithm~3}.

    \begin{algorithm}[t]
        \caption{Proposed block paritioning algorithm}\label{alg:proposed_grouping}
        {\small \begin{algorithmic}[1]
            \REQUIRE $\{\mu_i\}_{i=1}^K\text{ and } \epsilon$ 
            \STATE $\{R_i\}_{i=1}^K \gets {\bf Algorithm~2}(\{\mu_i\}_{i=1}^K,\epsilon)$
            \FOR {$p=1$ to $(R_{\max}-1)/2$}
            \STATE $R \gets R_{\max}-2(p-1)$
            \STATE $\mathcal{J}_p \gets \{j:R_j=R\}$
            \ENDFOR
            \STATE $\mathcal{J}_{\rm single} \gets \{j:R_j=1\}$
            \STATE $\mathcal{A} \gets \mathcal{J}_1$
            \STATE $g\gets 1$
            \FOR {$p=1$ to $(R_{\max}-1)/2-1$}
            \STATE $\mathcal{B}\gets \mathcal{J}_{p+1}$
            \STATE $\mu^{(1)}\gets \min_{i\in \mathcal{A}} \mu_i$
            \STATE $\mu^{(2)}\gets \min_{i\in \mathcal{B}} \mu_i$
            \STATE $\textsf{BLER}^{(1)} \gets 1-(1-\mu^{(1)})^{|\mathcal{A}|}$
            \STATE $\textsf{BLER}^{(2)} \gets 1-(1-\mu^{(2)})^{|\mathcal{B}|}$
            \STATE $Y\gets \frac{\sqrt{V}}{\ln 2}Q^{-1}(\textsf{BLER}^{(1)})$
            \STATE $\gamma_{\rm th} \gets {\rm Eq.}~\eqref{eq:C_star}$
            \IF {$Q^{-1}(\textsf{BLER}^{(1)})/Q^{-1}(\textsf{BLER}^{(2)})<\gamma_{\rm th}$}
            \STATE $\mathcal{A}\gets \mathcal{A}\cup \mathcal{B}$
            \STATE $\mathcal{K}_g \gets \mathcal{A}$
            \ELSE 
            \STATE $\mathcal{K}_g \gets \mathcal{A}$, $\mathcal{K}_{g+1} \gets \mathcal{B}$
            \STATE $g\gets g+1$
            \STATE $\mathcal{A} \gets \mathcal{J}_{p+1}$
            \ENDIF
            \ENDFOR
            \STATE $G\gets g$
            \ENSURE $\{\mathcal{K}_g\}_{g=1}^G \text{ and } \mathcal{J}_{\rm single}$
        \end{algorithmic}}
    \end{algorithm}

    \begin{algorithm}[t]
        \caption{Algorithm for fitting block sizes}\label{alg:size_fitting}
        {\small \begin{algorithmic}[1]
            \REQUIRE $\{\mathcal{K}_g\}_{g=1}^G, \mathcal{J}_{\rm single}\text{ and } \overline{\mathcal{K}}=\{\tilde{k}_1,\tilde{k}_2,\ldots,\tilde{k}_{|\overline{\mathcal{K}}|}\}$ 
            \STATE $i\gets 1$, $g\gets 1$, $K_{\rm block} = \sum_{g=1}^G| \mathcal{K}_g|$
            \STATE $k_{\rm acc} \gets 0$
            \WHILE {$k_{\rm acc}<K_{\rm block}$}
                \STATE $k^{\star}\gets \arg\min_{k\in\overline{\mathcal{K}}}||\mathcal{K}_g|-k|$
                \IF {{$|\mathcal{K}_g| < k^{\star}$ and $k_{\rm acc}+k^{\star} \leq K_{\rm block}$}}
                    \WHILE {$|\mathcal{K}_g| < k^{\star}$}
                        \STATE $\mathcal{K}_{g+1} \gets \mathcal{K}_g \cup \mathcal{K}_{g+1}$
                        \STATE $g\gets g+1$
                    \ENDWHILE
                \ENDIF

                \IF {$|\mathcal{K}_g| \geq  k^{\star}$}
                    \STATE $\Delta k \gets |\mathcal{K}_g|- k^{\star}$
                    \STATE $\Delta \mathcal{K}\gets \mathcal{K}_g[-\Delta k:\text{end}]$
                    \STATE $\mathcal{K}_i^{\prime} \gets \mathcal{K}_g \setminus \Delta \mathcal{K}$
                    \STATE $\mathcal{K}_{g+1}\gets \mathcal{K}_{g+1}\cup \Delta \mathcal{K}$
                    \STATE $k_{\rm acc}\gets k_{\rm acc}+ k^{\star}$
                \ELSIF {$|\mathcal{K}_g| < k^{\star}$ and $k_{\rm acc}+k^{\star} > K_{\rm block}$}
                    \IF{$k_{\rm acc}+k^{\star} > K$}
                    \STATE $\mathcal{K}_i^{\prime} \gets \{\mathcal{K}_g[1],\mathcal{K}_g[2],\ldots,K,0,\ldots,0\}$ so that $|K_i^{\prime}|=k^{\star}$
                    \STATE $\mathcal{J}_{\rm single}^{\prime} \gets \emptyset$
                    \ELSE 
                    \STATE $\mathcal{K}_i^{\prime} \gets \{\mathcal{K}_g[1],\mathcal{K}_g[2],\ldots,k_{\rm acc}+k^{\star}\}$ so that $|K_i^{\prime}|=k^{\star}$ 
                    \STATE $\mathcal{J}_{\rm single}^{\prime} \gets \{k_{\rm acc}+k^{\star}+1,k_{\rm acc}+k^{\star}+2,\ldots,K\}$
                    \ENDIF
                    \STATE $k_{\rm acc}\gets k_{\rm acc}+k^{\star}$
                \ENDIF
                \STATE $i\gets i+1$, $g\gets g+1$  
            \ENDWHILE
            \STATE $G^\prime \gets i-1$
            \ENSURE $\{\mathcal{K}_i^{\prime}\}_{i=1}^{G^\prime}\text{ and } \mathcal{J}_{\rm single}^{\prime}$ 
        \end{algorithmic}}
    \end{algorithm}

    \subsection{A Systematic Approach to Incorporating Modern Channel Codes}
    \begin{algorithm}[t]
        \caption{Algorithm for block-level UEP}\label{alg:LDPC_UEP}
        {\small \begin{algorithmic}[1]
            \REQUIRE $\{\mu_i\}_{i=1}^K$, $\epsilon$,  $\overline{\mathcal{K}}=\{\tilde{k}_1,\ldots,\tilde{k}_{|\overline{\mathcal{K}}|}\}$, $\mathcal{R}=\{\tilde{r}_1,\ldots,\tilde{r}_{|\mathcal{R}|}\}$, and $T(k,r)$ 
            \STATE $\{\{\mathcal{K}_g\}_{g=1}^G, \mathcal{J}_{\rm single} \} \gets {\bf Algorithm~3}(\{\mu_i\}_{i=1}^K,\epsilon)$
            \STATE $\{\{\mathcal{K}_g^\prime\}_{g=1}^{G^\prime}, \mathcal{J}_{\rm single}^{\prime} \} \gets {\bf Algorithm~4}(\{\mathcal{K}_g\}_{g=1}^G, \mathcal{J}_{\rm single}, \overline{\mathcal{K}})$
            \FOR {$g=1$ to $G^\prime$}
                \STATE $r_g \gets \max \; r\in\mathcal{R}$ s.t. $T(|\mathcal{K}_g^\prime|,r)< \min_{i \in \mathcal{K}_g^\prime} \mu_i$
            \ENDFOR
            \ENSURE $\{\mathcal{K}_g^\prime\}_{g=1}^{G^\prime}, \{r_g\}_{g=1}^{G^\prime} \text{ and } \mathcal{J}_{\rm single}^{\prime}$
        \end{algorithmic}}
    \end{algorithm}

    After determining the $G$ blocks $\{\mathcal{K}_g\}_{g=1}^G$, the semantic bits within each block can be jointly encoded by a channel encoder.
    To utilize the capacity-based analysis in \textbf{Proposition~2} and \textbf{Theorem~1}, which serve as theoretical guidelines under the assumption of capacity-achieving codes, we employ modern channel coding schemes to minimize the gap between theory and practice.
    Specifically, we employ polar codes and LDPC codes. Polar codes are known to achieve capacity at infinite blocklength and outperforming LDPC in shorter lengths \cite{van2016performance}, whereas LDPC codes typically exhibit better performance at moderate to high code rates and longer blocklength \cite{tahir2017ber}.  Therefore, we apply polar codes and LDPC codes for short and large blocklengths, respectively. Further details on adopted channel coding schemes are provided in Sec.~V.

    To ensure compatibility with practical polar/LDPC codes and reduce implementation complexity, we constrain the possible block sizes to a predefined set $\overline{\mathcal{K}}$. Note that larger block sizes tend to incur higher decoding complexity due to increased dimensions of the parity-check matrix and the iteration overhead in belief propagation decoding.
    Furthermore, to maintain tractable memory complexity, we restrict the possible code rates to the set $\mathcal{R}$, e.g. $\mathcal{R}=\{1/2, 1/3, 1/4\}$, motivated by modulation and coding scheme (MCS) tables commonly found in modern wireless standards \cite{3gpp38213}. Consequently, we modify the block sizes produced by \textbf{Algorithm~3} to align with the set $\overline{\mathcal{K}}$. This adjustment process, detailed in \textbf{Algorithm~4}, merges or splits adjacent blocks as necessary, ensuring each resulting block $\mathcal{K}_i^{\prime}$ closely matches an element of $\overline{\mathcal{K}}$. Specifically, if a block size is smaller than the required code length, subsequent blocks are merged to reach the necessary size. Conversely, if a block is excessively large, surplus bits are moved to the following block. 
    This ensures compatibility with practical polar and LDPC codebooks.

    To summarize the complete block-level UEP procedure, we present a unified algorithm incorporating the previously discussed block partitioning and rate-selection steps. As detailed in \textbf{Algorithm~5}, the procedure begins by executing the block partitioning algorithm (\textbf{Algorithm~3}), followed by adjustments to the block sizes to conform to the allowed block sizes (\textbf{Algorithm~4}). Subsequently, for each block, the most suitable code rate is selected from the predefined set $\mathcal{R}$ by referencing a lookup table $T(k,r)$ that maps block sizes and code rates to achievable bit-flip probabilities. The selected code rate ensures that the bit-flip probability of all bits within the block does not exceed the smallest TBP among its members. This integrated approach enables the block-level UEP framework to achieve varying protection levels systematically, while remaining fully compatible with practical codebooks.

    \section{Simulation Results}
    In this section, we evaluate the superiority of the proposed UEP frameworks for an image transmission task with the MNIST, CIFAR-10, and CIFAR-100 datasets \cite{MNIST_v2, CIFAR10}. The MNIST dataset contains 70,000 grayscale images of handwritten digits of size 28x28 pixels, divided into 60,000 training and 10,000 test samples. 
    The CIFAR-10 and CIFAR-100 datasets consist of 60,000 color images of size 32×32×3 pixels. CIFAR-10 is split into 50,000 training and 10,000 test samples across 10 classes, while CIFAR-100 includes 100 classes with the same train/test split.

\begin{table}[t]

        \renewcommand{\arraystretch}{1.1}

        {\caption{The semantic encoder and decoder architectures for MNIST, CIFAR-10, and CIFAR-100 datasets.}\label{table:autoencoder_architecture}}

        \setlength{\tabcolsep}{3pt}

        \scriptsize

        \centering


        {\begin{tabular}{|c|c|c|} \hline

            \multicolumn{2}{|c|}{} & \multicolumn{1}{c|}{Layers}   \\ \hline \hline

            \multirow{4}{*}{MNIST} & \multirow{2}{*}{Encoder} & \multicolumn{1}{c|}{C(32,3,1,2), PReLU, C(64,3,1,2), PReLU,}\\  
        & & \multicolumn{1}{c|}{C(64,5,2,1), PReLU, C(8,5,2,1)} \\ \cline{2-3} 
        & \multirow{2}{*}{Decoder} & \multicolumn{1}{c|}{CT(64,5,2,1,0), PReLU, CT(64,5,2,1,0), PReLU,} \\
        & & \multicolumn{1}{c|}{CT(32,3,1,2,0), PReLU, CT(1,4,1,2,0)} \\ \hline \hline
        
        & \multirow{2}{*}{Encoder} & \multicolumn{1}{c|}{C(64,5,2,2), PReLU, C(128,5,2,2), PReLU,} \\
        
                    CIFAR-10 & & \multicolumn{1}{c|}{C(128,5,1,2), PReLU, C(128,5,1,2), PReLU,C(24,5,1,2)}\\\cline{2-3} 
        
                    / CIFAR-100 & \multirow{2}{*}{Decoder} & \multicolumn{1}{c|}{CT(128,5,1,2,0), PReLU, CT(128,5,1,2,0), PReLU,} \\
        & & \multicolumn{1}{c|}{CT(128,5,1,2,0), PReLU, CT(64,5,2,2,1), PReLU,CT(3,5,2,2,1)} \\ \hline 
        

        \end{tabular}}\vspace{-2mm}


    \end{table}

    The encoder–decoder pair and the bit-flip probabilities $\{\mu_i\}$ are jointly trained as defined in \eqref{eq:loss_function}. We employ convolutional neural network (CNN)-based autoencoder architectures, following the design in \cite{bourtsoulatze2019deep}, which are detailed in Table~\ref{table:autoencoder_architecture}. In this table, ${\rm C}(c,k,s,p)$ represents a 2D convolutional layer with $c$ output channels, a $k\times k$ kernel, a stride of $s$, and padding of $p$. Similarly, ${\rm CT}(c,k,s,p,p_o)$ denotes a transposed 2D convolutional layer with output padding of $p_0$.
    The feature vector is quantized using 8 bits. The symbol transmission power is set to $P_{\rm trans} = 0$ dBW, and the received SNR defined as $|h|^2/2\sigma^2$ is fixed at 0 dB under the Rayleigh block fading channel.

    In the simulations, we compare the following channel coding frameworks:
    \begin{itemize}
        \item \textbf{Bit-UEP:} This is the proposed bit-level UEP framework. Based on the learned TBP $\{\mu_i\}$, we apply \textbf{Algorithm~2} to assign an individual repetition number to each bit. 
 
        \item \textbf{Block-UEP:} This is the proposed block-level UEP framework. Using the learned $\{\mu_i\}$, we apply \textbf{Algorithm~5} to perform block partitioning. The set of allowable block sizes is $\mathcal{K} = \{128,\,256,\,512,\,1024\}$. Each block is encoded using either a polar code (for block sizes of 128 and 256) or an LDPC code (for block sizes of 512 and 1024), where the code rate is selected from a lookup table with available code rates $\mathcal{R} = \{3/4,\,2/3,\,15/24,\,14/24,\,13/24,\,1/2,\,1/3\}$. Both polar and LDPC codes are implemented according to the 5G standard: polar codes utilize CRC-aided list decoding with a list size of 8, while LDPC codes are constructed using the standard's two base graphs and a lifting procedure \cite{3gpp38213}.

        \item \textbf{Repetition ($R_{\rm fix}$):} All bits are encoded using the repetition code with a fixed repetition number $R_{\rm fix}$.

        \item \textbf{LDPC:} This is an equal error protection scheme using an LDPC code. Block partitioning is performed identically to the \textbf{Block-UEP} scheme. However, all blocks are encoded using LDPC codes with a fixed rate, which is selected to ensure that the total blocklength matches that of the \textbf{Block-UEP} scheme.

        \item \textbf{Polar}: This is an equal error protection scheme using a polar code. Since a large block size is prohibitive due to polar code's implementation complexity, all semantic bit blocks are uniformly set to size 256. All blocks use the same code rate selected to ensure that the total blocklength matches that of the \textbf{Block-UEP} scheme.

        \item \textcolor{black}{\textbf{JSCC-UEP}: This baseline implements the joint source--channel UEP strategy in \cite{dong2017joint}, adapted to the semantic bitstream considered in our experiments. The ordered bitstream is divided into fixed equal-length groups, and each group is encoded using a polar code of length 256 with a code rate determined by its importance level.}

        \item \textcolor{black}{\textbf{BCH-UEP}: This baseline implements the semantic UEP scheme in \cite{xu2023unequal}. The semantic bits are divided into three fixed importance groups and protected by the $(63,36)$, $(63,45)$, and $(63,51)$ BCH codes, respectively.}
        
        \item \textbf{Genie:} This ideal baseline assumes perfect transmission with no bit-flips, i.e., all bits are received without error.
    \end{itemize}

    \begin{figure}
        \centering
        {\epsfig{file=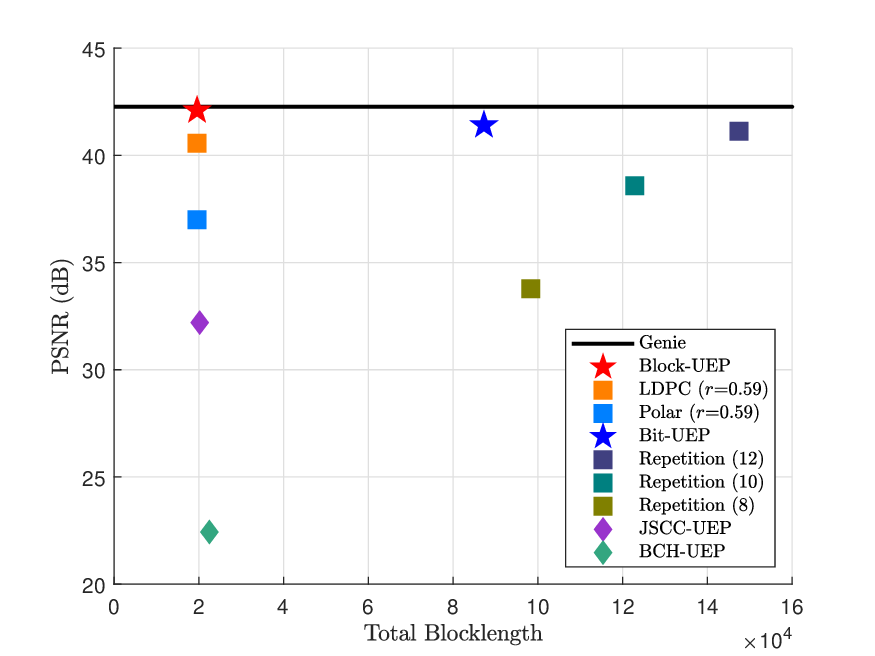,width=8cm}}\vspace{0mm}
        \caption{Comparison of PSNR versus total blocklength for various channel coding frameworks using the CIFAR-10 dataset with $\lambda=10^{-4}$ ($K=12288$).}  
        \label{fig:cifar_4_scatter}
    \end{figure}

    \begin{figure}
        \centering
        {\epsfig{file=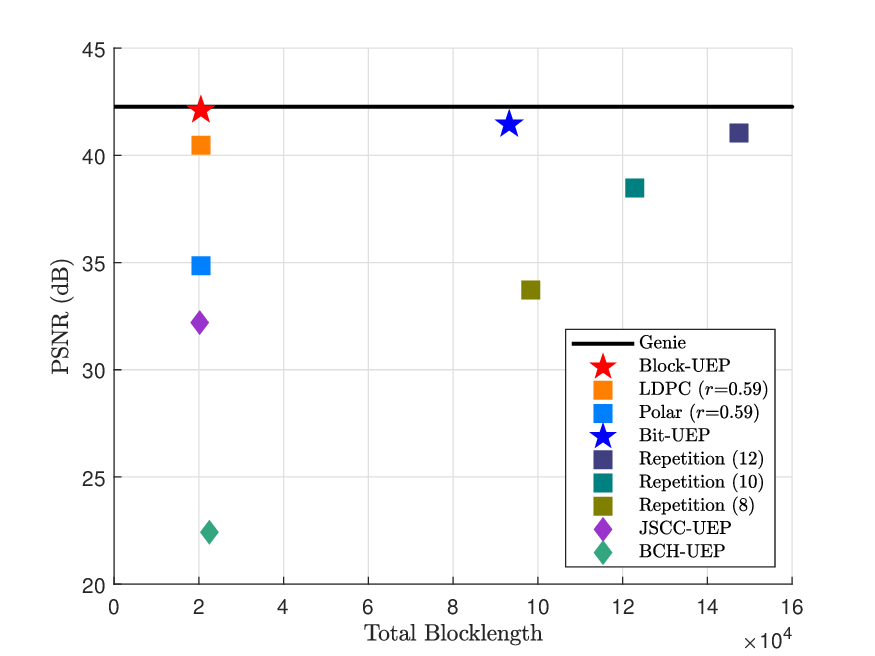,width=8cm}}\vspace{0mm}
        \caption{Comparison of PSNR versus total blocklength for various channel coding frameworks using the CIFAR-10 dataset with $\lambda=10^{-6}$ ($K=12288$).}  
        \label{fig:cifar_6_scatter}
    \end{figure}

    In Figs.~\ref{fig:cifar_4_scatter} and \ref{fig:cifar_6_scatter}, we compare the PSNR and corresponding total blocklength for various channel coding frameworks, using the CIFAR-10 dataset with $\lambda = 10^{-4}$ and $\lambda = 10^{-6}$, respectively. The results demonstrate that the proposed bit-level UEP framework (\textbf{Bit-UEP}) consistently achieves higher PSNR compared to fixed-rate repetition coding (\textbf{Repetition}) under both training settings. Specifically, the fixed-rate scheme with $R_{\rm fix}=8$ consumes slightly more blocklength than \textbf{Bit-UEP}, yet suffers from a significantly lower PSNR. On the other hand, the fixed-rate scheme with $R_{\rm fix}=12$ achieves comparable PSNR to \textbf{Bit-UEP} but at the cost of significantly larger blocklength usage. Moreover, coding gain is evident when comparing \textbf{Bit-UEP} and \textbf{Block-UEP}. \textbf{Block-UEP} achieves higher PSNR with even less blocklength usage compared to \textbf{Bit-UEP}. Additionally, \textbf{Block-UEP} outperforms traditional equal rate schemes (\textbf{LDPC} and \textbf{Polar}) in terms of PSNR while maintaining the same blocklength across both datasets. This result demonstrates the importance and effectiveness of dynamic partitioning of the semantic bits according to their protection requirements. 

\begin{figure}
        \centering
        {\epsfig{file=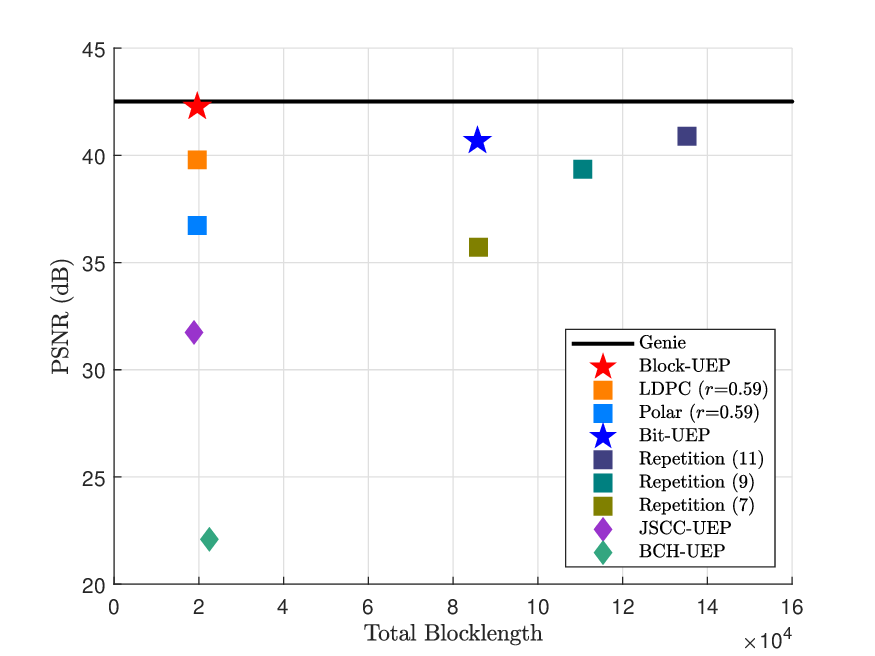,width=8cm}}\vspace{0mm}
        \caption{Comparison of PSNR versus total blocklength for various channel coding frameworks using the CIFAR-100 dataset with $\lambda=10^{-4}$ ($K=12288$).}  
        \label{fig:cifar100_4_scatter}
    \end{figure}

\begin{figure}
        \centering
        {\epsfig{file=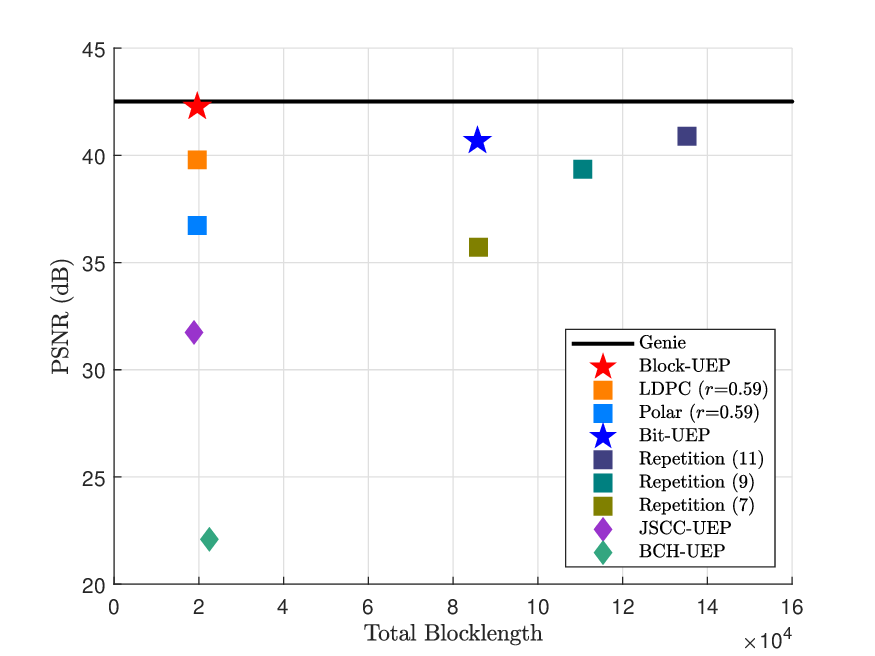,width=8cm}}\vspace{0mm}
        \caption{Comparison of PSNR versus total blocklength for various channel coding frameworks using the CIFAR-100 dataset with $\lambda=10^{-6}$ ($K=12288$).}  
        \label{fig:cifar100_6_scatter}
    \end{figure}

    Similarly, in Figs.~\ref{fig:cifar100_4_scatter} and \ref{fig:cifar100_6_scatter}, we compare the PSNR and total blocklength across various channel coding frameworks using the CIFAR-100 dataset, with $\lambda = 10^{-4}$ and $\lambda = 10^{-6}$, respectively. The results again confirm the superiority of unequal error protection over traditional equal protection schemes. From the consistent observations on both CIFAR-10 and CIFAR-100, we conclude that the proposed UEP frameworks provide clear advantages for semantic communication in terms of both reconstruction quality and transmission efficiency. Notably, the performance of \textbf{Block-UEP} approaches the genie-aided upper bound in both datasets, indicating its effectiveness in achieving near-optimal reconstruction quality with an improved transmission efficiency. In Fig.~\ref{fig:examples_scheme}, we  visualize this effect with reconstructed CIFAR-100 images. The figure demonstrates that \textbf{Block-UEP} offers the highest image quality, while the equal error protection  schemes (\textbf{LDPC} and \textbf{Polar}) suffer from image quality degradation.

    \textcolor{black}{The proposed \textbf{Block-UEP} also consistently achieves higher PSNR than the existing UEP baselines (\textbf{JSCC-UEP} and \textbf{BCH-UEP}) while using a comparable total blocklength. Both baseline schemes assign unequal protection levels based on predefined grouping structures. In contrast, the proposed algorithm adaptively groups bits according to the heterogeneous bit-level protection requirements. This result demonstrates the effectiveness of adaptive grouping for improving both reconstruction quality and transmission efficiency.}

    \textcolor{black}{To evaluate the generalizability of the proposed UEP framework to a different training methodology, we additionally adopt the information-theoretic training approach in \cite{oh2026towards}. This approach minimizes the reconstruction error subject to a mutual information constraint $I(\bm b;\hat{\bm b})\leq I_{\max}$ and optimizes the bit-flip probabilities through per-bit rate allocation, where $C_i=1-H_2(\mu_i)$, $H_2(\cdot)$ is the binary entropy function, and $\sum_i C_i=I_{\max}$. For this experiment, we set $I_{\max}=0.9K$. We then apply the proposed UEP algorithms to the resulting bit-flip probabilities without any modification.}

    \begin{figure}
        \centering
        {\epsfig{file=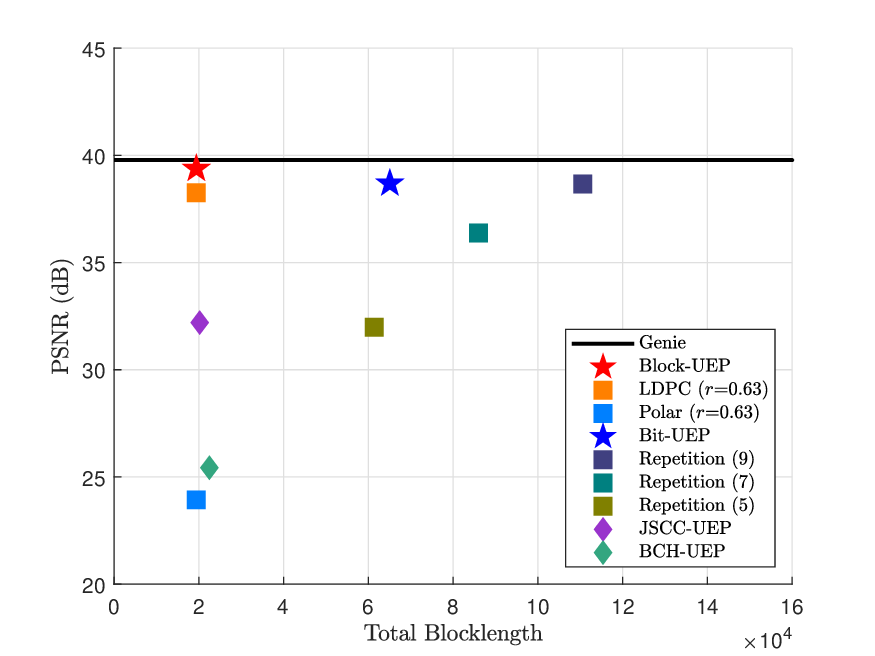,width=8cm}}\vspace{0mm}
        \caption{\textcolor{black}{Comparison of PSNR versus total blocklength for various channel coding frameworks using the CIFAR-10 dataset with bit-flip probabilities obtained from the information-theoretic training approach ($I_{\max}=0.9K$, $K=12288$).}}  
        \label{fig:cifar_Imax_scatter}
    \end{figure}

    \textcolor{black}{In Fig.~\ref{fig:cifar_Imax_scatter}, we compare the PSNR and total blocklength across various channel coding frameworks using the CIFAR-10 dataset when the  bit-flip probabilities obtained from the information-theoretic training approach with $I_{\max}=0.9K$ and $K=12288$. As shown in Fig.~\ref{fig:cifar_Imax_scatter}, the proposed UEP frameworks retain their performance advantages under this alternative training methodology. In particular, \textbf{Block-UEP} achieves higher PSNR than the equal-protection and existing UEP baselines while using a comparable total blocklength, and its performance approaches the genie-aided upper bound. This result demonstrates that the proposed framework can be applied to bit-flip probabilities learned using different training objectives and methodologies.}

    In Table~\ref{table:ordering}, we compare the PSNR of the proposed block-level UEP framework with different orderings of semantic bits. In the proposed ordering, the semantic bits are sorted in ascending order of their TBP, as described in Sec.~III and Sec.~IV, and \textbf{Algorithm~5} is applied to this sorted sequence. For comparison, we consider two alternative orderings: random and reverse. As shown in Table~\ref{table:ordering}, the proposed ordering achieves the highest PSNR, while the reverse ordering yields the worst performance. This result confirms that bits with lower TBP are more critical for image reconstruction. In Fig.~\ref{fig:ordering}, we also visualize this effect with reconstructed CIFAR-10 images. The figure demonstrates that reversing the bit order leads to severe degradation in image quality.

    \begin{table}
        \centering
        \caption{PSNR of the proposed block-level UEP framework with different orderings of semantic bits.}
        \begin{tabular}{ |c|ccc| } 
        \hline
        Dataset($\lambda$)& Proposed & Random & Reverse\\
        \hline
        CIFAR-10($10^{-6}$) & {\bf 42.24} & 29.33 & 20.00 \\
        \hline
        CIFAR-10($10^{-4}$) & {\bf 42.19} & 20.69 & 11.44 \\
        \hline
        MNIST($10^{-6}$) & {\bf 43.80} & 26.06 & 13.81 \\
        \hline
        MNIST($10^{-4}$) & {\bf 43.45} & 18.02 & 9.40\\
        \hline
        \end{tabular}
        \label{table:ordering}
    \end{table}

    \begin{figure}
        \centering
        {\epsfig{file=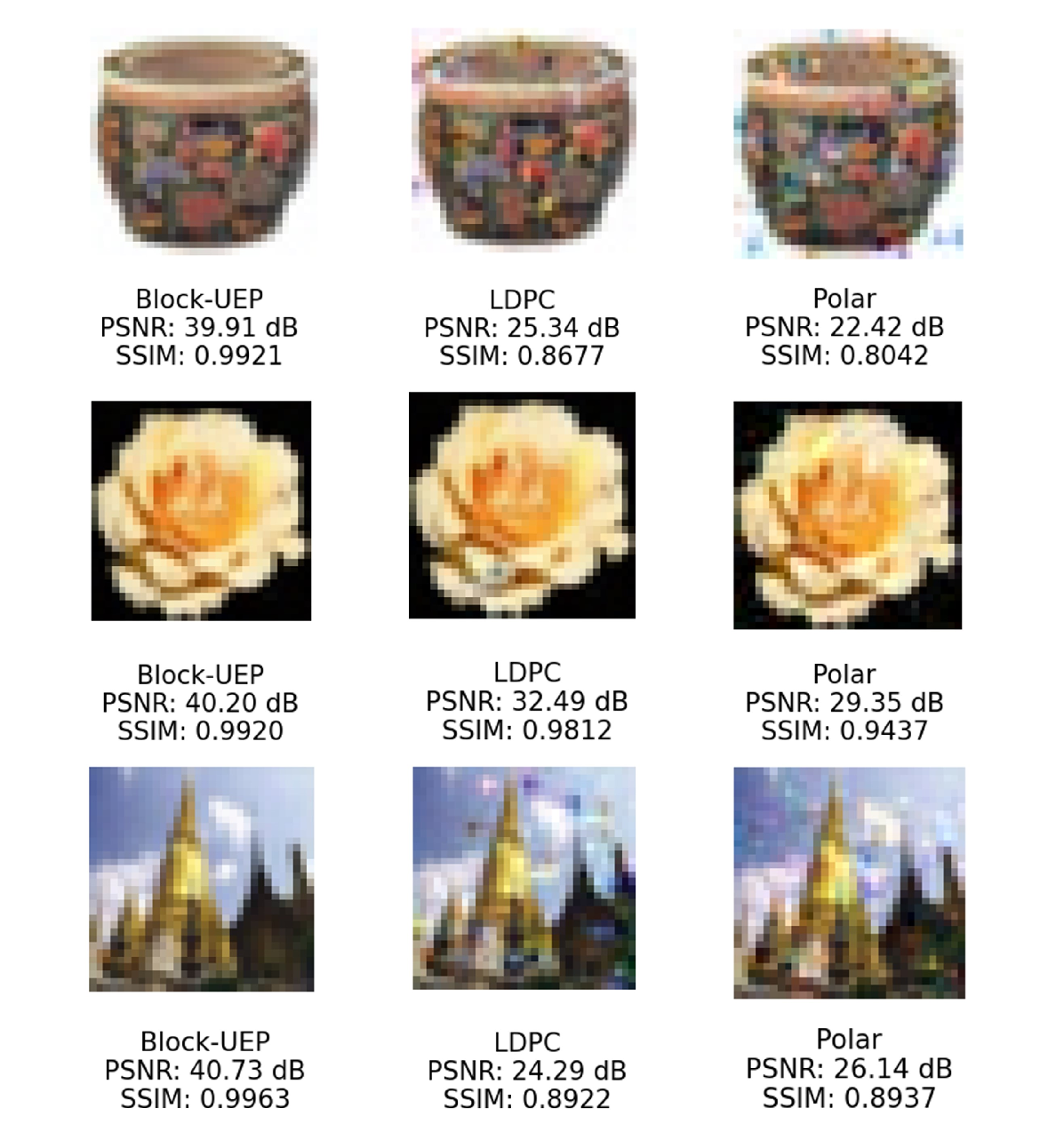,width=9cm}}\vspace{0mm}
        \caption{Visualization of reconstructed images and corresponding PSNR and SSIM values for various channel coding frameworks using CIFAR-100 dataset with $\lambda=10^{-4}$.}  
        \label{fig:examples_scheme}
    \end{figure}

    \begin{figure}
        \centering
        {\epsfig{file=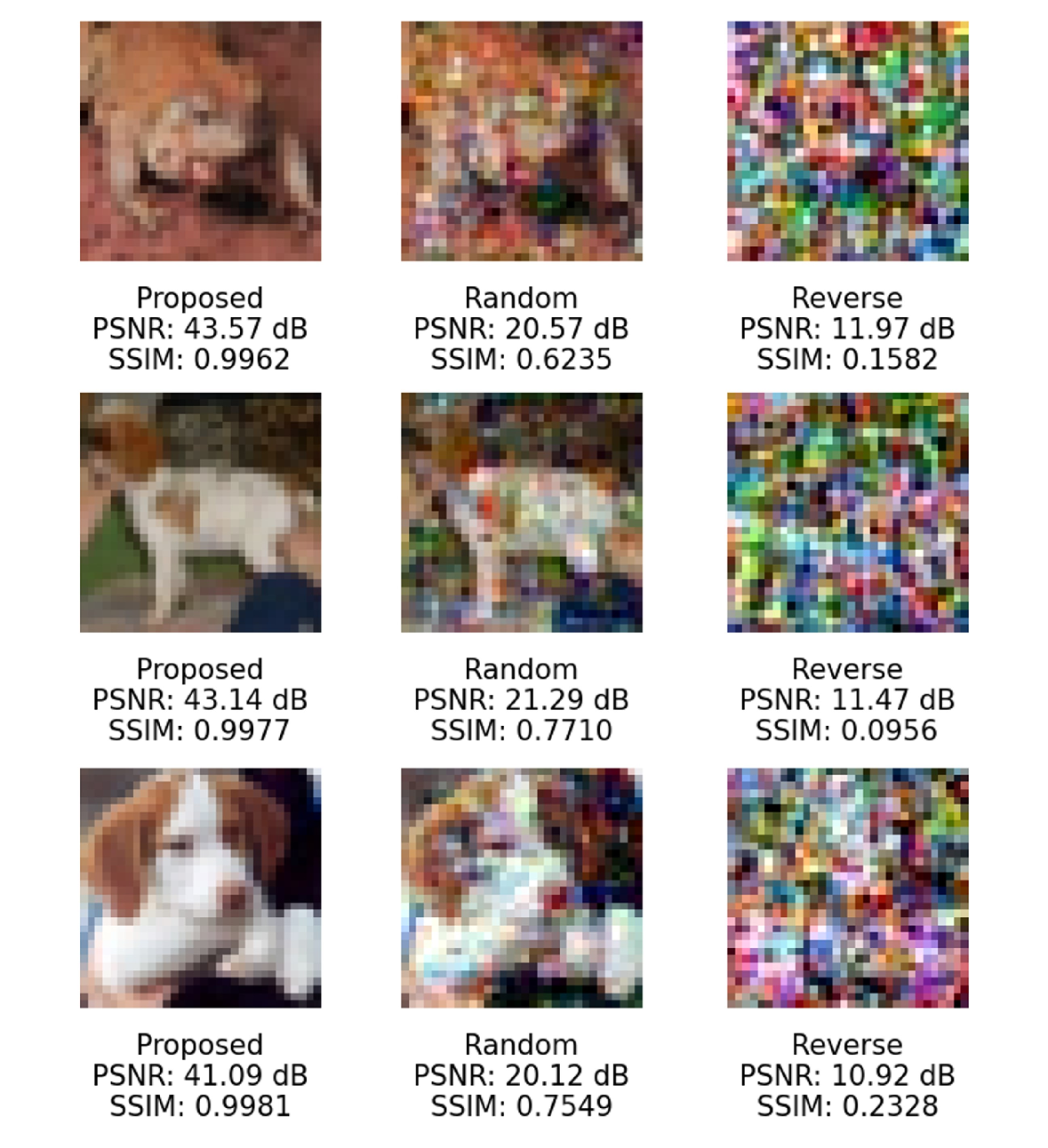,width=9cm}}\vspace{0mm}
        \caption{Visualization of reconstructed images and corresponding PSNR and SSIM values of the proposed block-level UEP framework with different orderings of semantic bits for $\lambda=10^{-4}$.}  
        \label{fig:ordering}
    \end{figure}

\begin{figure}
        \centering
        {\epsfig{file=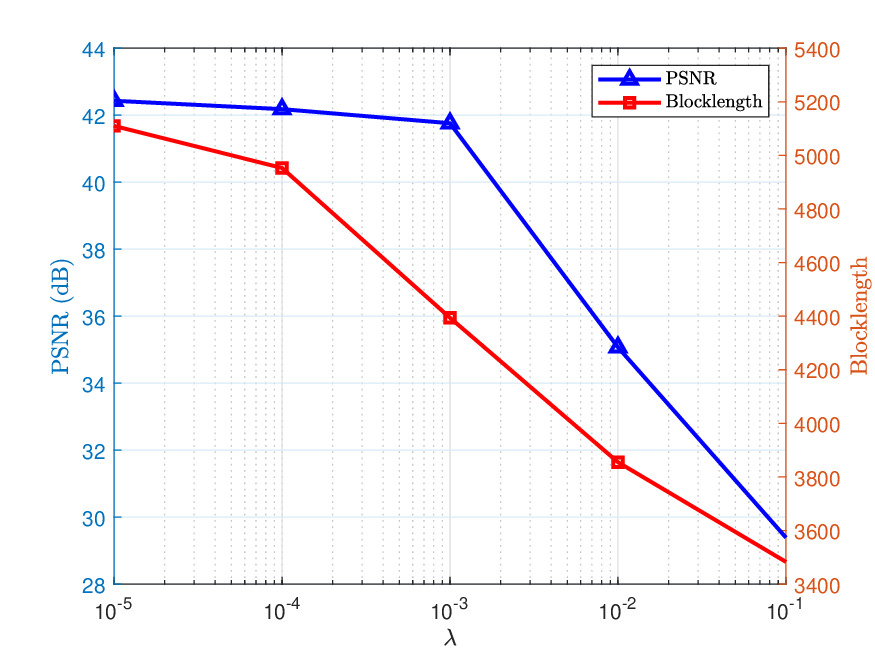,width=7cm}}\vspace{0mm}
        \caption{PSNR and total blocklength the proposed block-level UEP framework with various $\lambda$ values using the MNIST dataset ($K=3136$).}  
        \label{fig:mnist_lambda}
    \end{figure}

\begin{figure}
        \centering
        {\epsfig{file=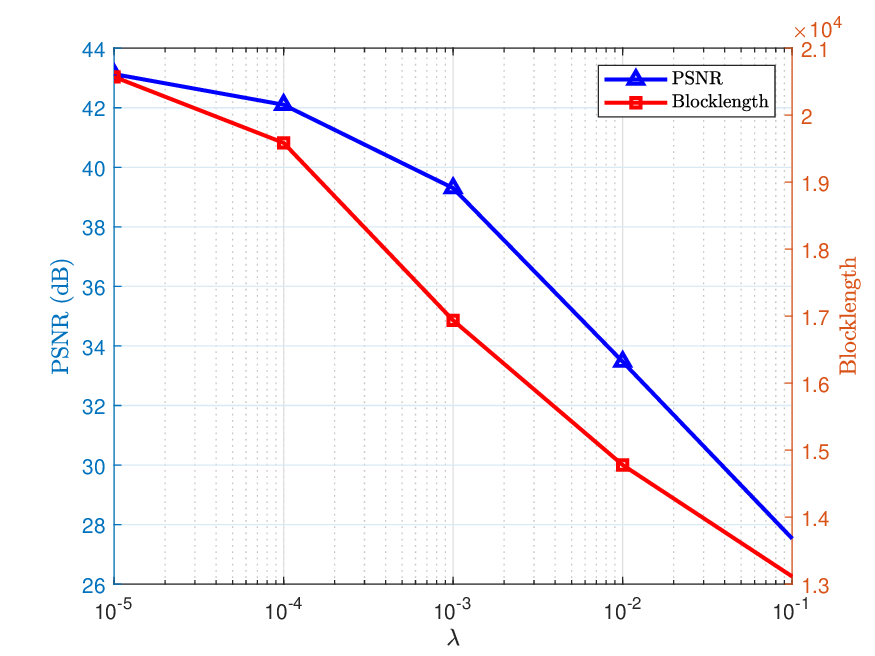,width=7cm}}\vspace{0mm}
        \caption{PSNR and total blocklength the proposed block-level UEP framework with various $\lambda$ values using the CIFAR-10 dataset ($K=12288$).}  
        \label{fig:cifar_lambda}
    \end{figure}

    In Figs.~\ref{fig:mnist_lambda} and \ref{fig:cifar_lambda}, we compare the PSNR and total blocklength of the proposed block-level UEP framework with different values of $\lambda$ using the MNIST and CIFAR-10 datasets, respectively. The regularization parameter $\lambda$ penalizes the assignment of excessively low bit-flip probabilities to task-irrelevant bits in the loss function \eqref{eq:loss_function}. As shown in the figures, a larger $\lambda$ tends to reduce PSNR, as it limits the model’s capacity to minimize reconstruction error. However, it also significantly decreases total blocklength in both datasets. These results suggest that selecting an appropriate value of $\lambda$ allows for a trade-off between reconstruction quality and transmission efficiency, depending on the desired task performance level.

    \section{Conclusion}
    In this paper, we explored the UEP problem in digital semantic communication by leveraging learned bit-flip probabilities as target protection levels. We proposed two coding frameworks: a bit-level UEP strategy based on repetition coding and a block-level UEP approach using polar codes and LDPC codes. The bit-level framework minimizes total blocklength while ensuring per-bit reliability, whereas the block-level framework enhances coding efficiency by partitioning bits with similar protection requirements into common blocks. For the block-level framework, we introduced a block partitioning algorithm guided by finite blocklength capacity analysis. Simulation results demonstrated that both frameworks significantly improve task performance and transmission efficiency compared to conventional methods. These results highlight the importance of bit-level error protection in semantic communication and lay the groundwork for future research on resource-efficient and reliability-aware semantic transmission strategies.

    A promising avenue for future research is to extend the proposed UEP framework to multi-user semantic communication scenarios, where users may experience heterogeneous channel conditions and demand different levels of quality-of-service (QoS). In such settings, efficient multiple access, user grouping and adaptive protection strategies become critical. Another important direction is to incorporate multi-antenna techniques to further improve transmission efficiency.

    \appendices
    \section{Proof of Proposition 1}\label{Apdx:Prop1}

    To show the boundedness by $r^{\rm (ub)}$, define a random variable $X$ which follows the binomial distribution ${\rm Bino}(R,\, \epsilon)$ so that $P_{\rm rep}(R)={\rm Pr}(X\geq R/2)$. Now, suppose that $\epsilon<1/2$ without loss of generality, and define $\delta = 1/(2\epsilon) -1$. Then, $P_{\rm rep}(R)$ is upper bounded as follows:
    \begin{align}
        P_{\rm rep}(R) &= {\rm Pr}\left(X \geq \frac{R}{2}\right)\\
        &={\rm Pr}(X \geq (1+\delta)\epsilon R)\\
        & \overset{(a)}{\leq} \exp(-R\cdot{\rm KL}((1+\delta)\epsilon||\epsilon))\\
        &=\exp\left(-R\cdot{\rm KL}\left(\frac{1}{2}||\epsilon\right)\right)\\
        &=\left(2\sqrt{\epsilon(1-\epsilon)}\right)^R,
    \end{align}
    where the inequality (a) is followed by Chernoff bound with Kullback–Leibler divergence \cite{chernoff_KL}.     \textcolor{black}{Let $a_\epsilon\triangleq2\sqrt{\epsilon(1-\epsilon)}$. Since $0<\epsilon<1/2$, we have $0<a_\epsilon<1$ and, consequently, $\ln a_\epsilon<0$. Therefore, the definition of $r^{\rm (ub)}$ gives}
    \begin{align}
        &\textcolor{black}{2r^{\rm (ub)}+1
        \geq\frac{\ln \mu}{\ln a_\epsilon}}\\
        &\textcolor{black}{\Rightarrow
        (2r^{\rm (ub)}+1)\ln a_\epsilon
        \leq \ln\mu}\\
        &\textcolor{black}{\Rightarrow \ln P_{\rm rep}(2r^{\rm (ub)}+1)
        \leq (2r^{\rm (ub)}+1)\ln a_\epsilon
        \leq \ln\mu}\\
        &\textcolor{black}{\Rightarrow P_{\rm rep}(2r^{\rm (ub)}+1)\leq \mu.}
    \end{align}

    \section{Proof of Proposition 2}\label{Apdx:Thm1}

    From $\textsf{BLER}(n_1,k_1)=\textsf{BLER}(n_2,k_2)$, we have
    \begin{align}
        Y\triangleq\sqrt{n_1}C-\frac{k_1}{\sqrt{n_1}}=\sqrt{n_2}C-\frac{k_2}{\sqrt{n_2}},
    \end{align}
    which gives $k_1=n_1C-Y\sqrt{n_1}$ and $k_2=n_2C-Y\sqrt{n_2}$, where $C=\log_2(1+\textsf{SNR})$. By plugging this into $k_1+k_2$, 
    \begin{align}
        &\quad  \sqrt{n_1+n_2}C-\frac{k_1+k_2}{\sqrt{n_1+n_2}}\\
        &= \sqrt{n_1+n_2}C-\frac{(n_1+n_2)C-Y(\sqrt{n_1}+\sqrt{n_2})}{\sqrt{n_1+n_2}}\\
        &=Y\frac{\sqrt{n_1}+\sqrt{n_2}}{\sqrt{n_1+n_2}}>Y.\\
    \end{align}
    Therefore, 
    \begin{align}
        &\quad  \textsf{BLER}(n_1+n_2,k_1+k_2)\\
        &=Q\left(\frac{\ln 2}{\sqrt{V}}\left(\sqrt{n_1+n_2}R-\frac{k_1+k_2}{\sqrt{n_1+n_2}}\right)\right)\\
        &<Q\left(\frac{\ln 2}{\sqrt{V}}Y \right)\\
        &=\textsf{BLER}(n_1,k_1).
    \end{align}
    By the monotonicity of $\textsf{BLER}(n,k)$ in $n$, there exists $n_3$ such that $\textsf{BLER}(n_3,k_1+k_2)=\textsf{BLER}(n_1,k_1)$ and $n_3<n_1+n_2$.
    
    \section{Proof of Theorem 1}\label{Apdx:Thm2}
    From $\textsf{BLER}(n_1,k_1)=\textsf{BLER}(n_3,k_1+k_2)$, we have
    \begin{subequations}
    \begin{align} 
        Y&=\sqrt{n_1}C-\frac{k_1}{\sqrt{n_1}} \label{eq:Y_1}\\
        &=\sqrt{n_3}C-\frac{k_1+k_2}{\sqrt{n_3}} \label{eq:Y_3}.
    \end{align}
    \end{subequations}
    Also, define $\gamma\triangleq Q^{-1}(\textsf{BLER}(n_1,k_1)) / Q^{-1}(\textsf{BLER}(n_2,k_2))$ so that measure the difference of BLER between $(n_1,k_1)$ code and $(n_2,k_2)$ code, which gives
    \begin{align} \label{eq:Y_2}
        Y=\gamma\left(\sqrt{n_2}C-\frac{k_2}{\sqrt{n_2}}\right).
    \end{align}
    Since \eqref{eq:Y_1}, \eqref{eq:Y_2} and \eqref{eq:Y_3} are all quadratic equation of $\sqrt{n_1}$, $\sqrt{n_2}$ and $\sqrt{n_3}$, respectively, quadratic formula gives
    \begin{align}
        \sqrt{n_1}&=\frac{1}{2C}\left(Y+\sqrt{Y^2+4k_1C} \right),\\
        \sqrt{n_2}&=\frac{1}{2C}\left(\frac{Y}{\gamma} +\sqrt{\frac{Y^2}{\gamma^2}+4k_2C} \right),\\
        \sqrt{n_3}&=\frac{1}{2C}\left(Y+\sqrt{Y^2+4(k_1+k_2)C} \right).
    \end{align}
    Therefore, 
    \begin{align}\label{eq:difference}
        &n_3-(n_1+n_2)=\frac{Y}{2C^2}\bigg(
        \sqrt{Y^2+4(k_1+k_2)C} \nonumber\\
        &\qquad \quad -\sqrt{Y^2+4k_1C}-\frac{1}{\gamma}\sqrt{\frac{Y^2}{\gamma^2}+4k_2C}-\frac{Y}{\gamma^2}
        \bigg).
    \end{align}
    From this relation, $n_3-(n_1+n_2)>0$ is equivalent to
    \begin{align}
        A-\frac{Y}{\gamma^2}>\frac{1}{\gamma}
        \sqrt{\frac{Y^2}{\gamma^2}+4k_2C},
    \end{align}
    where $A\triangleq\sqrt{Y^2+4(k_1+k_2)C}-\sqrt{Y^2+4k_1C}$. This inequality holds if and only if 
    \begin{align}
        A^2-\frac{2AY}{\gamma^2}-\frac{4k_2C}{\gamma^2}>0,
    \end{align}
    which is equivalent to
    \begin{align}
        \gamma>\gamma_{\rm th}(n_1,n_2,k_1,k_2).
    \end{align}
    This condition implies that merging two bit blocks whose BLERs sufficiently differ uses more blocklength.
    
    \bibliographystyle{IEEEtran}
    \bibliography{Reference}
    
\end{document}